\documentclass[12pt]{article}
\pdfoutput=1
\usepackage{amssymb,amsmath}
\usepackage{epsfig}
\usepackage{epstopdf}
\usepackage{latexsym}
\usepackage{graphicx}
\usepackage{subfigure}
\usepackage{booktabs}
\usepackage{bbm}
\usepackage{bbold}
\usepackage[margin=20pt,small]{caption}

\usepackage[toc]{appendix}

\usepackage{color}
\usepackage{datetime}

\ifpdf
\fi

\renewcommand{\theequation}{\arabic{section}.\arabic{equation}}

\def\cF{{\cal F}}

\def\cM{{\cal M}}
\def\cN{{\cal N}}
\def\cO{{\cal O}}
\def\cP{{\cal P}}

\def\cL{{\cal L}}

\def\cS{{\cal S}}

\def\cX{{\cal X}}
\def\cL{{\cal L}}
\def\cA{{\cal A}}

\def\cP{{\cal P}}

\def\cF{{\cal F}}
\def\cG{{\cal G}}
\def\cP{{\cal P}}
\def\cM{{\cal M}}
\def\cX{{\cal X}}
\def\cZ{{\cal Z}}

\definecolor{cardinal}{rgb}{0.6,0,0}
\definecolor{darkgreen}{rgb}{0,0.5,0}
\definecolor{golden}{rgb}{0.92, 0.7, 0}
\definecolor{midnight}{rgb}{0, 0, 0.5}
\definecolor{darkblue}{rgb}{0.2, 0, 0.8}

\newcommand{\be}{\begin{equation}}
\newcommand{\ee}{\end{equation}}
\newcommand{\bea}{\begin{eqnarray}}
\newcommand{\eea}{\end{eqnarray}}

\begin{document}

\begin{titlepage}


\bigskip
\bigskip
\bigskip
\centerline{\Large \bf Effective Temperature in Steady-state Dynamics}
\bigskip
\centerline{\Large \bf from Holography}
\bigskip
\bigskip
\centerline{{\bf Arnab Kundu}}
\bigskip

\bigskip
\centerline{Theory Division}
\centerline{Saha Institute of Nuclear Physics}
\centerline{1/AF Bidhannagar, Kolkata 700064, India.}
\bigskip
\bigskip
\bigskip
\centerline{arnab.kundu[at]saha.ac.in}
\bigskip
\bigskip

\begin{abstract}

\noindent  We argue that, within the realm of gauge-gravity duality, for a large class of systems in a steady-state there exists an effective thermodynamic description. This description comes equipped with an effective temperature and a free energy, but no well-defined notion of entropy. Such systems are described by probe degrees of freedom propagating in a much larger background, {\it e.g.}~$N_f$ number of $\cN =2$ hypermultiplets in $\cN=4$ $SU(N_c)$ super Yang-Mills theory, in the limit $N_f \ll N_c$. The steady-state is induced by exciting an external electric field that couples to the hypermultiplets and drives a constant current. With various stringy examples, we demonstrate that an {\it open string equivalence principle} determines a {\it unique} effective temperature for {\it all} fluctuations in the probe-sector. We further discuss various properties of the corresponding {\it open string metric} that determines the effective geometry which the probe degrees of freedom are coupled to. We also comment on the non-Abelian generalization, where the effective temperature depends on the corresponding sector of the fluctuation modes.  

\end{abstract}

\newpage

\tableofcontents

\end{titlepage}

\newpage

\section{Introduction}

Quantum Field Theory (QFT) provides us with an extremely rich framework in which we can formulate and analyze a microscopic description of various degrees of freedom in Nature, their interactions and emerging phenomenology. The standard treatment, such as through summing up Feynman diagrams {\it etc.} however, is completely perturbative and relies the existence of a suitable (dimensionless) {\it small} coupling in which the putative perturbation can be carried out.

Unfortunately, the existence of such a small coupling is far from guaranteed in physical systems. In fact, recent experimental endeavours have produced numerous examples of strongly-coupled systems across a wide range of energy-scales, such as the quark-gluon plasma (QGP) at the Relativistic Heavy Ion Collider (RHIC) and the Large Hadron Collider (LHC) at TeV-scale (see {\it e.g.}~\cite{Aamodt:2010pa}), or the cold atoms at unitarity at eV-scale (see {\it e.g.}~\cite{O'Hara:2002}). A plethora of strong-coupling physics data currently confronts us for a theoretical handle and understanding of the same from a first principle calculation.

Notwithstanding the non-perturbative formulations in QFT, the likes of Schwinger-Dyson equation or Ward-Takahashi identities that result from an underlying symmetry of the path integral, the so-called AdS/CFT correspondence\cite{Maldacena:1997re} has emerged to be an excellent tool for qualitative understanding of strong-coupling physics.\footnote{See {\it e.g.}~\cite{CasalderreySolana:2011us} for a review on some of these applications to particle physics, and \cite{Hartnoll:2009sz} for condensed matter-inspired systems.} The core of this correspondence is rooted in a more general gauge-string duality, for which a vast number of large $N_c$, $SU(N_c)$-type gauge theories can be explored.

Moreover, gauge-string duality turns out to be very crucial to analyze time-dependent (hence out-of-equilibrium) physics at strong coupling. Typically, we only have a handful of examples where such questions can be completely addressed within a QFT-framework, see {\it e.g.}~recent works in \cite{CalabreseCardy}. Moreover, many of these efforts rely heavily on conformal field theory in $(1+1)$-dimension, where an infinite dimensional conformal group strongly facilitates the analysis. Thus the governing principles and universal features, assuming they exist, in a time-dependent system remains elusive.

Time-dependent aspects in physics typically involve the physics of thermalization or a quench process, and there is already a vast literature on analysis from a holographic perspective. However, there is an intermediate state of a system, which is stationary but not static. This is normally known as the non-equilibrium steady-state (NESS). Such a state is described by a time-varying microscopic description, that yields a time-independent macroscopic observable.

We know that in equilibrium thermodynamics provides us with a remarkably successful macroscopic description of a system, in terms of a handful of extensive and intensive variables. It is therefore a natural question whether for a steady-state system an ``effective" or ``analogous" thermodynamic description exists. It is useful to note that an ``effective" thermal description of an inherently non-equilibrium system has already been observed in \cite{PhysRevLett.95.267001}-\cite{Karch:2010kt}, in systems at quantum criticality\cite{PhysRevLett.103.206401} or aging glass systems\cite{Cugliandolo.97}. More recently, far from equilibrium physics in a strongly coupled quantum critical system has been addressed in \cite{Bhaseen:2013ypa}.

In this article, which is a sequel to \cite{Kundu:2013eba}, we will use gauge-string duality to demonstrate that for a very large class of systems in a steady-state, an effective thermodynamic description holds. This effective thermodynamic description comes equipped with an effective temperature, {\it i.e.}~the intensive variable, but may not come with an ``effective entropy", {\it i.e.}~the canonically conjugate extensive variable.

\begin{figure}[h!]
\centering
\includegraphics[scale=0.45]{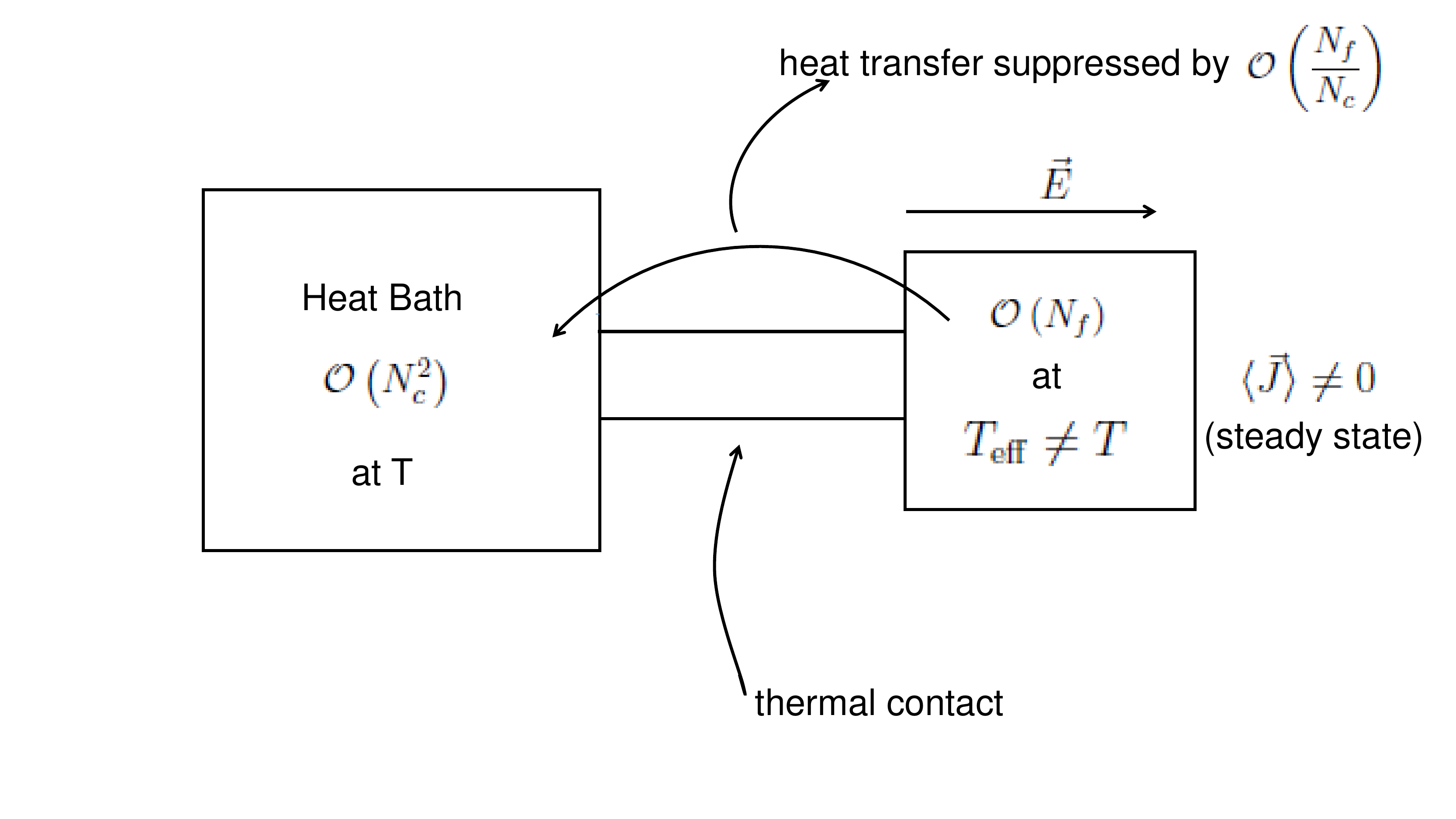}
\caption{\small A schematic diagram of our steady-state system. Here the $\cO(N_f)$ degrees of freedom are in a steady-state and coupled to a heat bath that consists of $\cO(N_c^2)$ degrees of freedom,}
\label{fig1}
\end{figure}

The prototype of our model is as follows: We consider a bath of a large number of degrees of freedom, and for the time being we will restrict ourselves to the adjoint sector of an  $SU(N_c)$-gauge theory in the large $N_c$ limit. In this heat bath, we want to introduce a ``probe" sector: typically this would be a fundamental matter sector with $N_f$ number of flavours. We will work in the limit $N_c \gg N_f$. In the dual gravitational description this is equivalent to studying the embeddings of a probe D$q$-brane in the background of a large number of D$p$-branes. The steady-state can be induced in the probe sector by exciting gauge fields on the probe worldvolume; this in turn corresponds to applying a constant electric field and inducing a steady current. This is schematically shown in Fig.~\ref{fig1}.

Thus we have a bath of $\cO(N_c^2)$ degrees of freedom kept at a temperature $T$ and we are introducing an $\cO(N_f)$ probe degrees of freedom which is in thermal contact with the bath. Now we excite a constant electric field in one of the gauge theory directions, which should drive a current through Schwinger pair production even at the absence of a charge density. Clearly, there will be a reverse energy-flux flowing from the probe sector to the heat bath that should eventually raise the temperature of the bath. However, in the limit $N_f \ll N_c$, this energy transfer is $N_f/N_c$ suppressed.

Note that, the D$p$-brane background are solutions of supergravity which essentially arises from the closed string sector. The dynamics of the probe D$q$-branes are described by the Dirac-Born-Infeld action which arises from the open string sector. We will argue, through various ``top-down" stringy constructions, that an open string equivalence principle determines the effective thermodynamics. 

This article is divided in the following parts: In section 2, we begin reviewing the dynamics of a probe D$7$-brane in the background of $N_c$ D$3$-branes. First we describe the classical physics and then move on to discuss the fluctuations on the probe. In this section we carefully derive all possible fluctuations of the D$7$-brane and argue that the fluctuations on the brane obey an open string equivalence principle and sense an ``effective" temperature. In section 3, we sketch out how to construct more examples following a similar approach. In section 4, we abstract away from the specific stringy construction and discuss general properties of the open string metric. In the next section we demonstrate that the same structure remains true for Schr\"{o}dinger-symmetric spacetime. In the following section we argue that a richer, but qualitatively similar physics emerges if we consider the non-Abelian generalization of the DBI-action. Finally we conclude in section 7. Some technical details have been relegated to two appendices.

\section{A model calculation: D3-D7 brane construction} \label{n4sym}

Let us begin by discussing the prototypical model for AdS/CFT, which is obtained by considering the near-horizon limit of a stack of $N_c$ D3-branes sitting at the tip of a singularity. Let us consider the singularity to be the tip of a cone whose base is a Sasaki-Einstein $5$-manifold, henceforth denoted by SE$_5$. When the SE$_5 \equiv S^5$, the $10$-dimensional geometry takes the form AdS$_5$-Schwarzschild$\times S^5$ and the dual field theory is the $\cN=4$ super Yang-Mills (SYM) theory with an $SU(N_c)$ gauge group.

The corresponding gravity background is given by\footnote{We are using the notation used in \cite{Albash:2007bq}.}
\begin{eqnarray} \label{metric1}
&& ds^2 = \frac{1}{4 r^2 R^2 } \left( - \frac{f^2}{\tilde{f}} dt^2 + \tilde{f} d\vec{x}^2 \right) + \frac{R^2}{r^2} dr^2 + R^2 d\Omega_5^2 \ , \\
&& f = 4 r^4 - b^4 \ , \quad \tilde{f} = 4 r^4 + b^4 \ , \\
&& F_{(5)} = \left(1 + \star \right) {\rm vol}_{S^5} = d C_{(4)} +  d \tilde{C}_{(4)} \ ,
\end{eqnarray}
where $t$ and $\vec{x} \equiv \{x^1, x^2, x^3\}$ are the field theory space-time directions, $r \in [b, \infty)$	 is the AdS-radial direction, $R$ is the curvature of AdS and $d\Omega_5^2$ is the metric on a unit $5$-sphere. The only matter field is a self-dual five-form $F_{(5)}$, for which the four-form potential is denoted by $C_{(4)}$ and $\tilde{C}_{(4)}$, whereas $\star$ denotes the Hodge dual operator. The unit sphere metric can be written as
\begin{eqnarray}
&& d\Omega_5^2 = d\theta^2 + \cos^2 \theta d\Omega_3^2 + \sin^2\theta d\phi^2 \ , \label{s3collapse} \\
&& d\Omega_3^2 = d\psi^2 + \cos^2 \psi d\beta^2 + \sin^2\psi d\gamma^2 \ . \label{s3met}
\end{eqnarray}
The parameter $b$ sets the radius of the black hole horizon and correspondingly the black hole temperature
\begin{eqnarray}
T = \frac{b}{\pi R^2} \ .
\end{eqnarray}
Evidently, the zero temperature limit is obtained by setting $b=0$. The radius of AdS sets the 't Hooft coupling for the dual field theory {\it via} $R^4 = \alpha'^2 g_{\rm YM}^2 N_c$, where $\alpha'$ is the string tension and $g_{\rm YM}$ is the gauge theory coupling.

The matter content of $\cN=4$ SYM is: the gauge field $A_\mu$, four adjoint fermions $\lambda$ and three complex scalars $\Phi^a$ ($a=1,2,3$). This theory has an $SU(4) \sim SO(6)$ R-symmetry, which corresponds to the rotational symmetry of the $S^5$ in the dual gravitational description. To introduce flavours, we need to introduce open string degrees of freedom which can be done by introducing probe D-branes of various dimensions along the lines of \cite{Karch:2002sh}. In particular, we can add $N_f$ probe D7-branes --- with the constraint $N_f \ll N_c$ to avoid any backreaction --- along the Poincare directions $\{t, \vec{x}\}$ and wrapping the $S^3 \subset S^5$. The D7-brane is a codimension $2$ object and therefore the profile can be specified by $\{\theta(r), \phi(r) \}$. Using the isometry along the $\phi$-direction, we can set $\phi(r)=0$ without any loss of generality and the probe brane profile will be entirely specified by $\theta(r)$.

Before proceeding further, let us offer some comments on the dual gauge theory side\footnote{In reviewing this part, we will closely follow the discussions in \cite{Myers:2007we}.}: adding D7-branes amounts to adding an $\cN=2$ hypermultiplets in the background of $\cN=4$ SYM. The hypermultiplets consists of two Weyl fermions, denoted by $\psi$ and $\bar{\psi}$ and two complex scalars, denoted by $q$ and $\bar{q}$. Now, $\{\psi, q\}$ transforms under the fundamental representation of SU$(N_c)$ and $\{\bar{\psi}, \bar{q}\}$ transforms under the anti-fundamental. The operators corresponding to the D7-brane profile functions $\theta$ and $\phi$ can be obtained to be
\begin{eqnarray}
&& \cos\theta \leftrightarrow \cO_m = i \bar{\psi} \psi + \bar{q} \left( m_q + \sqrt{2} \Phi^1\right) \bar{q}^{\dagger} + q^{\dagger} \left( m_q + \sqrt{2} \Phi^1\right) q + {\rm h.c.} \ , \\
&& \phi \leftrightarrow \cO_\phi = \bar{\psi} \psi + i \sqrt{2} \bar{q} \, \Phi^1 \bar{q}^{\dagger} + i \sqrt{2} q^{\dagger} \Phi^1 q + {\rm h.c.} \ ,
\end{eqnarray}
where $\Phi^1$ is a complex scalar field in the $\cN=4$ supermultiplet and $m_q$ is the mass of the fundamental quark.

The worldvolume theory is described by a $\cN=4$ SYM coupled to $N_f$ $\cN=2$ hypermultiplets. In the $\cN=1$ language, the vector multiplet of $\cN=4$ decomposes into the $\cN=1$ vector multiplet, denoted by $W_\alpha$, and $3$ chiral superfields, denoted by $\Phi_I$, with $I=1,2,3$. The $\cN=2$ matter sector can be written in terms of the $\cN=1$ chiral multiplets denoted by $Q^r$, $\tilde{Q}_r$, where $r = 1, \ldots, N_f$. The complete Lagrangian can be written in the $\cN=1$ notation as
\begin{eqnarray}
\cL & = & {\rm Im} \left[ \tau \int d^4\theta \left( {\rm tr} \left( \bar{\Phi}_I e^V \Phi_I e^{-V} \right) + Q_r^{\dagger} e^V Q^r  + \tilde{Q}_r^\dagger e^{-V} \tilde{Q}^r \right) \right. \nonumber\\  
 &+& \left. \tau \int d^2\theta \left( {\rm tr} \left(W^\alpha W_\alpha \right) + W \right) + {\rm c.c.} \right] \ ,
\end{eqnarray}
where
\begin{eqnarray}
W = {\rm tr} \left(\epsilon_{IJK} \Phi_i \Phi_j \Phi_K \right) + \tilde{Q}_r \left(m + \Phi_3 \right) Q^r \ .
\end{eqnarray}
More details on the degrees of freedom are summarized in table~\ref{table1}, where the $SO(4) \equiv SU(2)_\Phi \times SU(2)_R$ symmetry corresponds to the symmetry of the $S^3$ wrapped by the D7-brane. The transverse $SO(2)$ symmetry can be explicitly broken by separating the D7 from the D3-brane stack, which amounts to setting $m_q \not = 0$.
\begin{table}[ht] 
\centering  
\begin{tabular}{| c | c | c | c | c | c | c |} 
\hline
fields & components & spin & $SU(2)_\Phi \times SU(2)_R$ & $U(1)_R$ & $\Delta$ & $U(N_f)$ \\
\hline
$\Phi_1, \Phi_2$ & $X^4, X^5, X^6, X^7$ & $0$ & $\left(\frac{1}{2}, \frac{1}{2}\right)$ &  $0$ & $1$ & $1$  \\
 & $\lambda_1$, $\lambda_2$ & $\frac{1}{2}$ & $\left(\frac{1}{2}, 0 \right)$ & $-1$ & $\frac{3}{2}$ & $1$ \\
\hline
$\Phi_3$, $W_\alpha$ & $X_V^A = \left(X^8, X^9\right)$ & $0$ & $\left(0, 0 \right)$ & $+2$ & $1$ & $1$ \\
                 & $\lambda_3$, $\lambda_4$ & $\frac{1}{2}$ & $\left(0, \frac{1}{2} \right)$ & $+1$ & $1$ & $1$ \\
                 &  $v_\mu$ & $1$ & $\left(0, 0 \right)$ & $0$ & $1$ & $1$ \\
\hline
$Q$, $\tilde{Q}$ & $q^m = \left(q, \bar{q}\right)$ & $0$ & $\left(0, \frac{1}{2} \right)$  & $0$ & $1$ & $N_f$ \\
                          & $\psi_i = \left(\psi, \psi^\dagger\right)$  & $\frac{1}{2}$ &  $\left(0, 0 \right)$ & $\mp 1$ & $\frac{3}{2}$ & $N_f$ \\
\hline	                                         
\end{tabular} \caption{Degrees of freedom in the dual field theory. }  \label{table1}
\end{table} 

The action for the probe D7-brane is given by the Dirac-Born-Infeld (DBI) Lagrangian with an Wess-Zumino term\footnote{Here we are using the Lorentzian signature.}
\begin{eqnarray}
S_{\rm D7} & = & - N_f T_{{\rm D}7} \int d^8 \xi \, {\rm det}^{1/2} \left(P[G_{ab} + B_{ab}]  + 2\pi\alpha' f_{ab}\right)  \nonumber\\
& + & \left(2 \pi \alpha' \right)^2 \frac{\mu_7}{2} \int f_{(2)} \wedge f_{(2)} \wedge \left( P\left[C_{(4)} \right] + P \left[\tilde{C}_{(4)}\right]  \right)\ ,
\end{eqnarray}
where $P[G_{ab} + B_{ab}]$ denotes the pull-back of the NS-NS sector fields: $G$ denotes the closed-string background metric and B denotes the NS-NS field; $f_{ab}$ is the worldvolume $U(1)$ gauge field on the probe. The four-form potentials $C_{(4)}$ and $\tilde{C}_{(4)}$ yield the self-dual five-form. The collective coordinates $\{\xi\}$-denote D7-brane worldvolume coordinates, $T_{\rm D7}$ denotes the D7-brane tension and is given by $T_{\rm D7} = \mu_7/g_s$, where $g_s$ is the string coupling constant. Finally $N_f$ denotes the number of probe branes and we need to impose $N_f \ll N_c$ to suppress backreaction\footnote{For all practical purposes though, we will set $N_f=1$. This simplifies our analysis, otherwise we may need to worry about the non-Abelian generalization of the DBI-action. We will comment on this later. }.

To introduce a steady-state in the probe flavour sector, we will excite the gauge field\footnote{We use a convention in which all gauge fields on the worldvolume are denoted by small letters, and we absorb the factor of $(2\pi\alpha')$ in it.}
\begin{eqnarray} \label{gaugeansatz}
a_1 (r) = - E t + a(r) \ ,
\end{eqnarray}
where the subscript $1$ denotes the spatial direction $x^1$. Thus our ansatz (\ref{gaugeansatz}) breaks the $SO(3)$ rotational invariance down to $SO(2)$. The field $E$ corresponds to the field strength: $f_{1t} =  E$, which represents a constant electric field in the boundary theory. Note that, only the flavour degrees of freedom couple to this electric field. The function $a(r)$, which is hitherto undetermined, plays a more subtle role.

The resulting effective action for the probe D7-brane is given by
\begin{eqnarray} \label{lageff}
S_{\rm D7} = -T_{\rm D7} \int d^8\xi \frac{\cos^3\theta}{16 r^5} \left[ \tilde{f}^2 f^2 \left( 1+ r^2 \theta'^2 \right) -4 r^4 \left\{ -f^2\tilde{f} a'^2 + 4 \tilde{f}^2 R^4 E^2 \left( 1+ r^2 \theta'^2 \right) \right\} \right] ^{1/2} \ ,	
\end{eqnarray}
where we have set $(2\pi\alpha')=1$ for convenience and $' \equiv d / dr$. Let us further define the following dimensionless tilde-variables:
\begin{eqnarray}
r = b \tilde{r} \ , \quad a(r) = b \tilde{a}(r) \ , \quad E = \frac{b^2}{R^2} \tilde{E} \ , \quad \theta(r) = \theta(\tilde{r}) \quad {\rm with} \quad 2\pi\alpha'=1 \ .
\end{eqnarray}

The action in (\ref{lageff}) contains two dynamical functions: $\theta(r)$ and $a(r)$. The equation of motion for $a(r)$ immediately results in a first integral of motion
\begin{eqnarray} \label{eoma}
\tilde{a}' = - \frac{4 \tilde{J} \tilde{r} \sqrt{\tilde{g}}}{g} \sqrt{\frac{\left(16 \tilde{E}^2 \tilde{r}^4 - g\right) (1 + \tilde{r}^2 \theta'^2 ) }{64 \tilde{J}^2 \tilde{r}^6 - g^2 \tilde{g}^2 \cos^6\theta}} \ , \quad g = 4 \tilde{r}^4 -1 \ , \quad \tilde{g} = 4 \tilde{r}^4 + 1 \ .
\end{eqnarray}
where $\tilde{J}$ is an undetermined constant. The asymptotic behaviour of the function $a(r)$ can be obtained from (\ref{eoma})
\begin{eqnarray}
\lim_{\tilde{r} \to \infty} \tilde{a} (\tilde{r}) = 0 + \frac{\tilde{J}}{2 \tilde{r}^2} + \ldots \ ,
\end{eqnarray}
where the first term, which is the non-normalizable term in the near-boundary expansion of the function $\tilde{a}$, is set to zero by hand. Using the derivation in \cite{Karch:2007pd}, the constant $\tilde{J}$ is related to the vacuum expectation value of the flavour current which is induced by the applied electric field
\begin{eqnarray}
\langle J^1 \rangle \equiv \langle \bar{\psi} \gamma^1 \psi \rangle = - (2 \pi^2) b^3 V_{\mathbb R^3}  T_{\rm D7} \tilde{J} \ , 
\end{eqnarray}
where $V_{\mathbb R^3}$ represents the volume of the ${\mathbb R^3}$.\footnote{Here we have used that $2\pi\alpha'=1$ and $N_f=1$. Reinstating these factors, the relation becomes: $\langle J^1 \rangle =  - (4 \pi^3 \alpha') N_f b^3 V_{\mathbb R^3}  T_{\rm D7} \tilde{J}$.}

Thus the physical meaning of the function $a(r)$ becomes clear: it encodes the response current in the flavour sector, which is sourced by the applied constant electric field. We will momentarily come back to how this current is determined in terms of the applied electric field and the temperature of the background. Before discussing further, let us offer a few comments over the physical meaning of the profile function $\theta(r)$. From the action (\ref{lageff}), it is clear that the equation of motion for $\theta$ is rather involved. Let us instead focus on the large $\tilde{r}$ behaviour, where the equation of motion becomes
\begin{eqnarray}
\frac{d}{d\tilde{r}} \left( \tilde{r}^5 \theta' \right) + 3 \tilde{r}^3 \theta = 0 \ , 
\end{eqnarray}
which yields
\begin{eqnarray}
\theta (\tilde{r}) = \frac{\tilde{m}}{\tilde{r}} + \frac{\tilde{c}}{\tilde{r}^3} + \ldots \ , 
\end{eqnarray}
where $\tilde{m}$ and $\tilde{c}$ are two constants, which are subsequently identified as the bare quark mass and the quark condensate {\it via}\cite{Babington:2003vm, Mateos:2006nu, Albash:2006ew}
\begin{eqnarray}
m_q = \frac{b \tilde{m}}{2\pi\alpha'} \ , \quad \langle \bar{\psi} \psi \rangle  = - 8 \pi^3 \alpha' V_{\mathbb R^3} N_f T_{\rm D7} b^3 \tilde{c} \ .
\end{eqnarray}
In the absence of any external electric field, this flavour sector undergoes a topology changing first order phase transition at finite temperature\cite{Mateos:2006nu, Albash:2006ew}. The phase transition is driven by tuning the dimensionless ratio $(m_q/T)$, where $T$ is the background temperature. The corresponding two phases --- which in the dual theory describes a phase of bound mesons and a plasma one --- are characterized by qualitatively different embeddings of the D7-brane: the so called Minkowski embedding and the black hole embedding, respectively. For more details on this, the reader is referred to \cite{Mateos:2006nu, Albash:2006ew}.

Now we will discuss how the boundary current is determined dynamically. Let us consider eliminating the function $\tilde{a}$ in favour of the constant $\tilde{J}$. This can be achieved by considering the following Legendre transformation 
\begin{eqnarray} \label{actionleg}
I_{\rm D7} & = & S_{\rm D7} - \int d^8\xi \frac{\partial \cL_{\rm D7}}{\partial F_{r1}} F_{r1} = - 2 \pi^2 V_{\mathbb R^3} N_f T_{\rm D7} b^4 \tilde{I}_{\rm D7} \ , \\
\tilde{I}_{\rm D7} &=& \int dt d\tilde{r} \left[ \frac{\sqrt{1+\tilde{r}^2 \theta'^2}}{16 \tilde{r}^5 g \sqrt{\tilde{g}}} \sqrt{\left(g^{2} - 16 \tilde{r}^4 \tilde{E}^2 \right) \left( -64 \tilde{r}^6 \tilde{g}^2 \tilde{J}^2 + g^{2} \tilde{g}^3 \cos^6 \theta\right)} \right] \ ,
\end{eqnarray}
Now, on physical grounds, we must require $\tilde{I}_{\rm D7}$ remains real. Thus, the two factors inside the square root must change sign at the same location, denoted by $\tilde{r}_*$ and this results in the following algebraic conditions
\begin{eqnarray}
\left( 4 \tilde{r}_*^4 -1 \right)^2 - 16 \tilde{r}_*^4 \tilde{E}^2 = 0 \quad \implies \quad \tilde{r}_*^2 = \frac{\tilde{E}+ \sqrt{\tilde{E}^2 + 1 }}{2} \ ,
\end{eqnarray}
and subsequently
\begin{eqnarray} \label{ohm1}
\tilde{J}^2 = \frac{\left( 4 \tilde{r}_*^4 -1 \right)^2 \left( 4 \tilde{r}_*^4 + 1 \right)^3 \cos^6\theta\left(\tilde{r}_*\right)}{64 \tilde{r}_*^6 \left( 4 \tilde{r}_*^4 + 1 \right)^2} \ .
\end{eqnarray}
Thus we have finally obtained an Ohm's law with a conductivity which is a non-linear function of the electric field itself. The location $r_*$, which is above the location of the background event horizon, plays a crucial role in the dynamics. We will henceforth call this a {\it pseudo horizon}.

A few comments are in order. Note that the current vanishes when $\tilde{r}_*^4 = 1/4$ or $\theta = \pi/2$. The first case corresponds to having a vanishing electric field, where the pseudo horizon and the event horizon coincide. The second scenario corresponds to the case when the $S^3$, wrapped by the D7-brane, shrinks to a zero size as observed from (\ref{s3collapse}). The shrinking $S^3$ corresponds to the Minkowski embeddings and thus corresponds to the bound meson states in the dual field theory. Since there is no charge carrier in this phase, the current is identically zero.

For simplicity, and this will suffice for our purposes, let us consider the case when the black hole in the bulk geometry disappears, {\it i.e.}~$b=0$. The action in this case is given by
\begin{eqnarray} \label{actionbzero}
S_{\rm D7} = - T_7 \int d^8 \xi r^3 \cos^3\theta \left[ \left(1 - \frac{R^4 E^2}{r^4}\right) \left(1 + r^2 \theta'^2\right) + a'^2 \right]^{1/2} \ .
\end{eqnarray}
From (\ref{actionbzero}), the pragmatic role of the function $a(r)$ becomes clear: If we set $a(r) = {\rm const}$, then the action vanishes at $r_*^2 = R^2 E$, which is the location of the pseudo horizon. However, there is nothing special about this point in the bulk and hence the probe branes cannot end there. By exciting a non-zero $a'$, we are allowed to go beyond the pseudo horizon.

The boundary conditions for the profile function $\theta$ can be fixed following \cite{Karch:2006bv}. The pulled-back D7-brane metric contains the following terms
\begin{eqnarray}
\frac{dr^2}{r^2} \left(1 + r^2 \theta'^2\right) + \cos^2\theta d\Omega_3^2  \subset ds_{\rm D7}^2 \ ,
\end{eqnarray}
which leads to an excess angle in the $\{\cos\theta, \Omega_3\}$-plane. The coefficient of this excess angle is proportional to $r^{-2} \theta'^{-2}$ evaluated at $r=r_{\rm min}$, where $r_{\rm min}$ denotes the minimum radius where the Minkowski embedding reaches. Evidently, $r_{\rm min} \ge r_{*}$. To kill off this excess angle, we need to impose 
\begin{eqnarray} \label{bcmin}
\left. \theta \right|_{r_{\rm min}} = \pi/2 \ , \quad \left. \theta' \right|_{r_{\rm min}} = - \infty \ . 
\end{eqnarray}
Finally the boundary conditions for the {\it pseudo horizon embeddings} can be fixed from the equation of motion 
\begin{eqnarray} \label{bcbh}
\left. \theta \right|_{r_{*}} = \theta_0 \ , \quad \left. \theta' \right|_{r_*} = - \frac{1}{\hat{r}_*} \tan\frac{\theta_0}{2} \ ,
\end{eqnarray}
where the hatted variable is defined as $r = R \sqrt{E} \hat{r}$. Clearly, the qualitative difference between the Minkowski embedding and the {\it pseudo horizon embedding} is how the brane ends, {\it i.e.}~without or with a conical singularity.

We remind the reader that as far as the qualitative behaviour of the probe embedding is concerned, there are characteristic similarities to the physics in a thermal background which was analyzed in \cite{Mateos:2006nu, Albash:2006ew} including the first order phase transition. The pseudo horizon plays a role qualitatively similar to a background event horizon in this respect. We will further argue that this feature is rather robust. In the next section we will study the fluctuations in the bosonic sector.

\subsection{Fluctuations: bosonic sector}

Let us study the fluctuation modes on the probe brane. These fluctuation modes will correspond to the meson operators in the dual $\cN=2$ field theory. For example, the scalar meson operators are given by\footnote{In reviewing this section, we will follow the notations and discussions in \cite{Kirsch:2006he}.}
\begin{eqnarray}
\cO_{\rm scalar}^{A \ell} = \bar{\psi}_i \sigma_{ij}^A \, X^\ell \psi_j + \bar{q}^m X_V^A X^\ell q^m \ , \quad i , m = 1,2 \ ,
\end{eqnarray}
with conformal dimensions $\Delta = 3 + \ell$. Here $\sigma^A = \left(\sigma_1, \sigma_2\right)$ denotes the Pauli matrices doublet, and $X_V^A$, $q^m$, $\psi_i$ are defined in table \ref{table1}. $X^\ell$ denotes the symmetric traceless operator $X^{\{ i_1 \ldots i_\ell \}}$ of $\ell$ adjoint scalars $X^i$, for $i = 4,5,6,7$. The mesonic operators, $\cO_{\rm scalar}^{A\ell}$ transform in the $\left(\frac{\ell}{2}, \frac{\ell}{2}\right)$ of the $SO(4)$ and have $+2$ charge under the $SO(2) \equiv U(1)_R$.

To obtain the fluctuation action, we need to expand the following probe action up to quadratic order\footnote{From here on, we will use $\delta\phi$ as the fluctuation corresponding to $\phi$ on the worldvolume, where $\phi$ is a generic field defined on the probe.}
\begin{eqnarray}
S_{\rm probe} & = & - T_{\rm D7} \int d^8\xi \sqrt{-{\rm det} E} + \mu_7 \int_{\cM_8} \delta f_{(2)} \wedge f_{(2)} \wedge P\left[\tilde{C}_{(4)} \right] \nonumber\\
&+& \frac{1}{2} \mu_7 \int_{\cM_8} \delta f_{(2)} \wedge \delta f_{(2)} \wedge P\left[C_{(4)}\right] \ ,
\end{eqnarray}
where
\begin{eqnarray}
&& E = P[G + B] + f \ ,  \quad {\rm and \, \, with}  \quad r^2 = \frac{1}{2} \left( u^2 + \sqrt{u^4 - b^4}\right) \\
&& C_{(4)} = \frac{1}{g_s} \frac{u^4}{R^4} \, dt \wedge dx^1 \wedge dx^2 \wedge dx^3 \ , \\
&& \tilde{C}_{(4)} = - \frac{R^4}{g_s} \left( 1 - \cos^4\theta\right) \sin\psi \cos\psi \, d\psi  \wedge d\alpha \wedge d\beta \wedge d\phi \ . 
\end{eqnarray}
The fluctuations can be represented by
\begin{eqnarray}
&& \theta = \theta^{(0)} + \delta\theta \ , \quad \phi = \phi^{(0)} + \delta\phi \ , \\
&& f_{ab} = f_{ab}^{(0)} + \delta f_{ab} \ ,
\end{eqnarray}
where $\theta^{(0)}$, $\phi^{(0)}$ and $f_{ab}^{(0)}$ are the classical fields and $\delta\theta(\xi^a)$, $\delta\phi(\xi^a)$, $\delta f_{ab}(\xi^a)$ denote the scalar and vector fluctuations. The fluctuation action is then obtained to be
\begin{eqnarray} \label{scalaraction}
S_{\rm scalar} & = & - T_{\rm D7} \int d^8\xi \, \sqrt{- {\rm det} E^{(0)}} \, \left[\frac{1}{2} \cS^{ab} g_{ab}^{(2)} + \right. \nonumber\\
&& \left.\frac{1}{2} \left( \frac{1}{2} \cS^{aa'} \cS^{bb'} + \frac{1}{4} \cS^{a'b} \cS^{b'a} + \cS^{aa'} \cA^{bb'} + \cA^{aa'} \cA^{bb'} \right) g_{a'b}^{(1)} \, g_{ab'}^{(1)}\right] \ , \\
S_{\rm vector} & = &  - T_{\rm D7} \int d^8\xi \, \sqrt{- {\rm det} E^{(0)}} \, \frac{1}{4} \left(\cS^{aa'} \cS^{bb'} + \cA^{aa'} \cA^{bb'} + \frac{1}{2} \cA^{a'b} \cA^{b'a} \right) \left(\delta f_{a'b}\right) \left(\delta f_{b'a} \right) \nonumber\\
& + & T_{\rm D7} \int_{\cM_8} \frac{u^4}{2R^4} \, \delta f_{(2)} \wedge \delta f_{(2)} \wedge dt \wedge dx^1 \wedge dx^2 \wedge dx^3 \ ,
\end{eqnarray}
and a coupled term
\begin{eqnarray} \label{coupledaction}
S_{\rm coupled} & = & - T_{\rm D7} \int d^8\xi \, \sqrt{- {\rm det} E^{(0)}} \, \frac{1}{2} \left( \cS^{aa'} \cA^{bb'} + \cS^{bb'} \cA^{aa'} + \frac{1}{2} \cS^{a'b} \cA^{ab'} \right) g_{a'b}^{(1)} \left(\delta f_{b'a}\right) \nonumber\\
& - & \frac{T_{\rm D7}} {R^4} \left( 1 - \cos^4 \theta_0\right) \int_{\cM_8} \sin(2\psi)  \left(\partial_a \delta\phi \right) \delta f_{(2)} \wedge f_{(2)} \wedge d\psi \wedge d\alpha \wedge d\beta \wedge d\xi^a \ ,
\end{eqnarray}
where we have defined
\begin{eqnarray}
g_{ab}^{(2)} & = & G_{\theta\theta} \left(\partial_a \delta\theta\right) \left(\partial_b \delta\theta\right) + G_{\phi\phi} \left(\partial_a \delta\phi\right) \left(\partial_b \delta\phi\right) + \frac{1}{2} \left(\partial_\theta^2 G_{mn} \right)\delta_a^m \delta_b^n \left(\delta\theta\right)^2 \ , \\
g_{ab}^{(1)} & = & G_{\theta\theta} \, \theta'(r) \left[ \left(\partial_a \delta \theta\right) \delta_{br} + \left(\partial_b \delta \theta\right) \delta_{ar} \right] + \left(\partial_\theta G_{mn} \right) \delta_{a}^m \delta_b^n \left(\delta\theta \right) \ , \label{gab11}
\end{eqnarray}
where $\delta_{ar}$, $\delta_{br}$ etc denote the Kronecker delta; $G_{\theta\theta} = R^2 $ and $G_{\phi\phi} = R^2 \sin^2\theta$ denotes the corresponding metric components in the $10$-dimensional geometry in (\ref{metric1}). The metric $G_{mn}$ represents the one on the $S^3$ defined in (\ref{s3collapse}) and (\ref{s3met}), where $m,n = 1, 2, 3$. The quantities $\cS$ and $\cA$, which are the crucial objects, are defined in (\ref{defSA1})-(\ref{defSA3}). Further note that the actions in (\ref{scalaraction})-(\ref{coupledaction}) are written in complete generality, {\it i.e.}~without making any assumptions about the form of $f_{ab}$ and therefore about $\cS$ and $\cA$. Evidently, the bosonic fluctuations in (\ref{scalaraction})-(\ref{coupledaction}) are complicated.

We will now focus on simple cases when the modes decouple. A remarkable simplification occurs if we focus on the $\theta_0(r) = 0$ embedding, which corresponds to classically introducing massless flavours. Furthermore, if we concentrate on fluctuations oscillating only along the non-compact directions, then from (\ref{gab11}) we conclude that $g_{ab}^{(1)} = 0$ and thus $S_{\rm coupled} = 0$ identically. Furthermore, by noting that the only non-vanishing components of $\cA$ are $\cA^{tx} = - \cA^{xt}$ and $\cA^{xu} = - \cA^{ux}$, it can be checked that 
\begin{eqnarray}
\left(\cA^{aa'} \cA^{bb'} + \frac{1}{2} \cA^{a'b} \cA^{b'a} \right) \left(\delta f_{a'b}\right) \left(\delta f_{b'a} \right) = 0 \ .
\end{eqnarray}
With these simplifications, the effective scalar and vector fluctuation actions take very simple forms
\begin{eqnarray}
S_{\rm scalar} & = & - T_{\rm D7} \int d^8\xi \, \sqrt{- {\rm det} E^{(0)}} \, \frac{1}{2} \cS^{ab} \left[ G_{\theta\theta} \left(\partial_a \delta\theta\right) \left(\partial_b \delta\theta\right)  \right] \ , \label{scasim1} \\
S_{\rm vector} & = & - T_{\rm D7} \int d^8\xi \, \sqrt{- {\rm det} E^{(0)}} \, \frac{1}{4} \cS^{aa'} \cS^{bb'}  \left(\delta f_{a'b}\right) \left(\delta f_{b'a} \right) \ . \label{vecsim1}
\end{eqnarray}
The fluctuation for $\delta \phi$ is multiplied by an overall factor of $\sin^2\theta_0$, which vanishes for the massless case; hence this mode is absent from the above effective actions.\footnote{It can be checked rigorously that the equations of motion resulting from (\ref{scalaraction})-(\ref{coupledaction}) do admit this consistent truncation.} The effective action describes a free scalar and a massless vector mode propagating in a background geometry which is governed by $\cS$, the so-called open string metric.

Let us now explicitly write down the elements of $\cS$ and $\cA$ including the gauge field in (\ref{gaugeansatz}). They can be represented as
\begin{eqnarray}
\cS = \cS_{tx^1 u} \otimes \cS_{x^2x^3} \otimes \cS_{S^3} \ , \quad \cA = \cA_{tx^1 u} \otimes \cA_{x^2x^3} \otimes \cA_{S^3} \ ,
\end{eqnarray}
where $\cS_{x^2x^3}$ is identical to the metric components in that plane, $\cS_{S^3}$ is identical to the metric components along the $S^3$ and $\cA_{x^2x^3} = 0$, $\cA_{S^3} = 0$. The non-triviality arises in the $\{t, x^1\equiv x, u\}$-plane. The non-vanishing components are
\begin{eqnarray}
&& \cS_{tt} = G_{tt} + \frac{E^2}{G_{xx}} \ , \quad \cS_{tu} = \frac{E a'}{G_{xx}} = \cS_{ut} \ ,  \label{sym1} \\
&&  \cS_{xx} = \frac{E^2}{G_{tt}} + G_{xx} + \frac{a'^2}{g_{uu}} \ , \quad \cS_{uu} = g_{uu} + \frac{a'^2}{G_{xx}} \ , \label{sym2}
\end{eqnarray}
where
\begin{eqnarray}
g_{uu} = G_{uu} + \theta'^2 G_{\theta\theta} \ ,
\end{eqnarray}
where $'$ now denotes taking derivative with respect to $u$. The non-vanishing components of the anti-symmetric tensor $\cA$ are given by
\begin{eqnarray}
&& \cA^{tx} =  - \frac{ E g_{uu}}{g_{uu} \left(G_{tt} G_{xx} + E^2\right) + G_{tt} a'^2} = - \cA^{xt} \ , \label{asym1} \\
&& \cA^{xu} =   \frac{ a' G_{tt}}{g_{uu} \left(G_{tt} G_{xx} + E^2\right) + G_{tt} a'^2} = - \cA^{ux} \ . \label{asym2}
\end{eqnarray}
We also note that the gauge field is given by
\begin{eqnarray} \label{gd3d7}
a'(u) = J \left(\frac{G_{tt}G_{xx} +E^2}{G_{tt}} \right)^{1/2} \left( \frac{G_{uu} + \theta'^2 G_{\theta\theta}}{J^2 + G_{tt} G_{xx}^2 \cos^2(\theta)}\right)^{1/2} \ .
\end{eqnarray}
Let us further investigate the properties of $\cS$. It is clear from the form of the fluctuation action, $\cS$ plays the role of an effective metric as far as the fluctuations are concerned.\footnote{This fact can be explicitly checked by focussing on the simplest case, where one sets $\theta_0 = 0$ and considers only modes which oscillate along the time direction, {\it i.e.}~$\delta X \sim \delta X(u) e^{- i \omega t} $. Here $\delta X$ denotes a generic fluctuation field and $\omega$ denotes its frequency.} Therefore the effective geometry in the $\{t, x, u\}$-plane is described by a metric\footnote{It is evident that the ``effective" metric is identical to the pull back metric on the worldvolume of the probe in all other directions, hence we are focussing only in this subspace.}
\begin{eqnarray}
ds_{\rm eff}^2 & = & \cS_{tt} dt^2 + 2 \cS_{ut} du dt + \cS_{uu} du^2 + \cS_{xx} dx^2 \nonumber\\
& = & \cS_{tt} d\tau^2 + \left(\cS_{uu} - \frac{\cS_{ut}^2}{\cS_{tt}}\right) du^2 + \cS_{xx} dx^2 \ ,
\end{eqnarray}
where we have redefined the time coordinate
\begin{eqnarray}
t = \tau + h(u) \ , \quad {\rm where} \quad h'(u) = \frac{\cS_{ut}}{\cS_{tt}} \ .
\end{eqnarray}
Rewriting the $10$-dimensional background as
\begin{eqnarray}
&& ds_{10}^2 = - \frac{u^2}{R^2} f(u) dt^2 + \frac{u^2}{R^2} dx^2 + \frac{u^2}{R^2} \left(dx_2^2 + dx_3^2\right) + \frac{R^2}{u^2 f(u)} du^2 + R^2 d\Omega_5^2 \ , \\
&& f(u) = 1 - \frac{u_H^4}{u^4} \ ,
\end{eqnarray}
we get
\begin{eqnarray} \label{effmet1}
ds_{\rm osm}^2 & = & - \frac{u^2}{R^2} \left( f - \frac{E^2R^4}{u^4} \right) d\tau^2 + \gamma_{uu} du^2 + \gamma_{xx} dx^2 + \frac{u^2}{R^2} dx_\perp^2  + R^2 \cos^2\theta d\Omega_3^2 \ ,
\end{eqnarray}
with
\begin{eqnarray}
&& f(u) = 1 - \frac{u_H^4}{u^4} \ , \quad dx_\perp^2 = dx_2^2 + dx_3^2 \ , \\
&& \gamma_{uu} = \frac{R^2 \left(\theta '^2 \left(u^6 f \cos ^6\theta +J^2 L^6\right)+u^4 \cos ^6\theta  \right)}{u^6 f \cos ^6 \theta - J^2 R^6} \ , \\
&& \gamma_{xx} = \frac{\left(E^2 R^4-u^4 f\right) \left(\theta '^2 \left(u^6 f \cos ^6 \theta + J^2 R^6\right) + u^4 \cos ^6 \theta \right)}{R^2 \left(u^2 f \theta '^2+1\right) \left(J^2 R^6 - u^6 f \cos ^6 \theta \right)} \ .
\end{eqnarray}
The geometry described by (\ref{effmet1}) has an event horizon at
\begin{eqnarray} \label{phori2}
\left. f - \frac{E^2R^4}{u^4} \right |_{u=u_*} = 0 \ ,
\end{eqnarray}
where $u_*$ is precisely the location of the pseudo-horizon. Furthermore, the constant $J$ can be fixed by demanding that the coefficient in front of $dx^2$ remains positive, which will be equivalent to setting $(J^2 R^6 - u^6 f \cos^2 \theta)=0$ where (\ref{phori2}) holds.\footnote{Note that, this purely geometric condition determines the flavour current in the boundary theory and is a special case of the open string metric membrane paradigm discussed in \cite{Kim:2011qh}.}

\subsection{Fluctuations: fermionic sector}

Let us now discuss the fermionic fluctuations, which correspond to the supersymmetric partners of the mesonic operators. These fermionic meson operators are of two types\footnote{Here also we will follow the notations used in \cite{Kirsch:2006he}, where the zero temperature fermionic meson spectrum has been worked out in details.}
\begin{eqnarray}
&& \cF_\alpha^\ell \sim  \bar{q} X^\ell \tilde{\psi}_\alpha^\dagger + \tilde{\psi}_\alpha X^\ell  q \ , \\
&& \cG_\alpha^\ell \sim \bar{\psi}_i \sigma_{ij}^A \lambda_{\alpha B} X^\ell \psi_j + \bar{q}^m X_V^A \lambda_{\alpha B} X^\ell q^m \ , \quad  A, B = 1, 2 \ ,
\end{eqnarray}
with conformal dimensions $\Delta = \frac{5}{2} + \ell$ and $\Delta = \frac{9}{2} + \ell$. The fermionic part of the D7-brane action at the quadratic order is given by\footnote{The general form of the quadratic action for a D$p$-brane is given in appendix C in {\it e.g.}~(\ref{actferm}) and (\ref{actferm2}).}
\begin{eqnarray} \label{fermd3d7}
S_{\rm spinor} = \frac{T_{\rm D7}}{2} \int d^8\xi \sqrt{-{\rm det} E^{(0)}} \, \bar{\Psi} \cP_{-}  \left(\tilde{E}_{(0)}^{-1}\right)^{ab} \Gamma_b D_a \Psi \ , \quad \tilde{E}_{ab} = g_{ab}^{(0)} + \tilde{\Gamma}_{(10)} f_{ab} \ ,
\end{eqnarray}
where $\xi^a$ corresponds to the worldvolume coordinates, which are identified with spacetime coordinates $\{t, x^1, x^2, x^3, u, \psi, \beta, \gamma \}$; $\Psi$ represents a doublet Majorana-Weyl spinor in $10$-dimensions
\begin{eqnarray} \label{doublespinor1}
\Psi =
\left( {\begin{array}{c}
\Psi_1  \\
\Psi_2 
\end{array} } \right) \ ,
\end{eqnarray}
where $\Psi_{1,2}$ each has positive chirality. In appendix B, we have defined $\tilde{\Gamma}_{(10)}$. The worldvolume gamma matrices $\Gamma_a$ are the pull back of the $10$-dimensional gamma matrices: $\Gamma_a = \left(\partial_a X^\mu\right) \Gamma_\mu $, where $a = 0 , \ldots, 7$ and $\mu = 0, \ldots , 9$. The operator $\cP_{-}$ is a kappa symmetry projection operator: $\cP_{-} = 1 - \Gamma_7$, where $\Gamma_p$ has been defined in (\ref{gammaiib}). In particular, we have
\begin{eqnarray}
\Gamma_7 & = & \frac{i \sigma_2}{\sqrt{- {\rm det} E^{(0)}}} \left[ \frac{\sigma_3}{2 \cdot 6!} \epsilon^{a_1 a_2 b_1 \ldots b_6} f_{a_1 a_2} \Gamma_{b_1\ldots b_6} - \frac{1}{8 \cdot 4!} \epsilon^{a_1 \ldots a_4 b_1 \ldots b_4} f_{a_1 a_2} f_{a_3 a_4} \Gamma_{b_1\ldots b_4} \right. \nonumber\\
&+& \left. \frac{\sigma_3}{16 \cdot 3!} \epsilon^{a_1\ldots a_6 b_1b_2} f_{a_1a_2} f_{a_3a_4} f_{a_5a_6} \Gamma_{b_1b_2} \right] \ ,
\end{eqnarray}
where $f_{ab}$ is the classical gauge field on the worldvolume of the probe.\footnote{The above expression is written for a general gauge field on the worldvolume of the probe; but for the specific ansatz in (\ref{gaugeansatz}), the last two terms above vanishes identically. Written explicitly, the operator $\cP_{-}$ in this case takes the following form
\begin{eqnarray}
\cP_{-} = 1 - \frac{i \sigma_2}{\sqrt{- {\rm det} E^{(0)}}} \sigma_3 \Gamma_{x^2} \Gamma_{x^3} \left[- E \left(\Gamma_{r} + \theta' \Gamma_{\theta} \right) - a' \Gamma_t \right] \Gamma_{\psi} \Gamma_{\beta} \Gamma_{\gamma} \ .
\end{eqnarray}
}

Let us now closely follow the discussion in \cite{Martucci:2005rb}, where one proceeds by fixing the $\kappa$-symmetry on the worldvolume of the probe. One can use the covariant gauge-fixing $\tilde{\Gamma}_{(10)} \Psi = \Psi$, {\it i.e.}~set $\Psi_2=0$, and this will yield the following action for the fermionic fluctuations
\begin{eqnarray} \label{ferm3}
S_{\rm spinor} & = & \frac{T_{\rm D7}}{2} \int d^8\xi \sqrt{-{\rm det} E^{(0)}}  \left[ \bar{\Psi}_1 \left(E_{(0)}^{-1}\right)^{ab} \Gamma_b D_a \Psi_1 - \bar{\Psi}_1 \hat{\Gamma}_{7}^{-1} \left(E_{(0)}^{-1}\right)^{ab} \Gamma_b W_a \Psi_1  \right]  \ ,
\end{eqnarray}
where
\begin{eqnarray}
\hat{\Gamma}_p = \left(-1\right)^{\frac{(p-2)(p-3)}{2}} \frac{1}{(p+1)!} \frac{\epsilon^{a_1 \ldots a_p}} {\sqrt{-{\rm det} E^{(0)}}} \sum_q \frac{\Gamma^{a_1\ldots a_{2q}}}{q! 2^q} f_{a_1a_2} \ldots f_{a_{2q-1}a_{2q}} \ ,
\end{eqnarray}
and $W_a$ is defined in (\ref{wiib}). We can now interpret the fermionic fluctuation action in (\ref{ferm3}) as a Dirac action with an effective mass term where the background geometry is again dictated by the $(g^{(0)}+f)^{-1}$, which is precisely the combination that appeared for the bosonic fluctuations. The action in (\ref{ferm3}) can be further simplified. Note that the gamma matrices $\Gamma_\mu$ generate the Clifford algebra associated with the worldvolume induced metric 
\begin{eqnarray}
\left \{\Gamma_a, \Gamma_b \right \} = 2 g_{ab}^{(0)} = 2 P[G]_{ab} \ ,
\end{eqnarray}
where $P[G] $ stands for the pull back of the background geometry. Let us now define
\begin{eqnarray}
\gamma^a = \left(E_{(0)}^{-1}\right)^{ab} \Gamma_b \ , 
\end{eqnarray}
which then yields
\begin{eqnarray} \label{osmgamma}
\left\{\gamma^a , \gamma^b\right\} = 2 \cS^{ab} \ .
\end{eqnarray}
With this, the fermionic action now becomes
\begin{eqnarray} \label{ferm4}
S_{\rm spinor} & = & \frac{T_{\rm D7}}{2} \int d^8\xi \sqrt{-{\rm det} E^{(0)}}  \left[ \bar{\Psi}_1 \gamma^a D_a \Psi_1 - \cM \bar{\Psi}_1  \Psi_1  \right]  \ ,
\end{eqnarray}
which results in the following equation of motion 
\begin{eqnarray}
\gamma^a D_a \Psi_1 - \cM \Psi_1 = 0 \ , \quad \cM = \hat{\Gamma}_7^{-1} \gamma^a W_a  \ .
\end{eqnarray}
The above equation is not analytically solvable, so we will not discuss its solutions. However, it is evident that the effective geometry as measured by the open string metric determines the propagation of the fermionic fluctuations as well, thus demonstrating the existence of an equivalence principle in the open string sector. From this and the open string metric given in (\ref{effmet1}) it is easy to check that on the trivial embedding $\theta_0 =0$ we need to solve the Dirac equation in the open string metric along $\{t, \vec{x}, u\}$-directions with an effective mass term.\footnote{The modes along the internal directions will contribute to this effective mass.}

\subsection{Some properties of the open string metric}

In the previous two sections we have explicitly demonstrated that all the fluctuation modes on the probe experience the open string metric geometry, therefore an {\it open string equivalence principle} holds for these modes. 
In this section, we will further investigate some properties of this metric. Let us begin by writing the open string metric once more
\begin{eqnarray} \label{osm1}
ds_{\rm osm}^2 & = & - \frac{u^2}{R^2} \left( f - \frac{E^2R^4}{u^4} \right) d\tau^2 + \gamma_{uu} du^2 + \gamma_{xx} dx^2 + \frac{u^2}{R^2} dx_\perp^2  + R^2 \cos^2\theta d\Omega_3^2 \ , 
\end{eqnarray}
with
\begin{eqnarray}
&& f(u) = 1 - \frac{u_H^4}{u^4} \ , \quad dx_\perp^2 = dx_2^2 + dx_3^2 \ , \\
&& \gamma_{uu} = \frac{R^2 \left(\theta '^2 \left(u^6 f \cos ^6\theta +J^2 R^6\right)+u^4 \cos ^6\theta  \right)}{u^6 f \cos ^6 \theta - J^2 R^6} \ , \\
&& \gamma_{xx} = \frac{\left(E^2 R^4-u^4 f\right) \left(\theta '^2 \left(u^6 f \cos ^6 \theta + J^2 R^6\right) + u^4 \cos ^6 \theta \right)}{R^2 \left(u^2 f \theta '^2+1\right) \left(J^2 R^6 - u^6 f \cos ^6 \theta \right)} \ .
\end{eqnarray}
The metric in (\ref{osm1}) is asymptotically AdS, and in the infrared (obtained by solving $f - E^2 R^4/u^4 = 0$) it behaves like an AdS-BH background. The geometry in (\ref{osm1}), however, has a curvature varying with the radial coordinate $u$ with a singularity located at $u=0$. Note that this singularity is present even when the closed string background does not have a black hole inside, {\it i.e.}~in the case when $u_H=0$.

For further discussion, let us focus on the case $\theta_0 =0$ (assuming $u_H=0$): the trivial embedding. In this case, we can forget about the $S^3$-directions and analyze the effective $5$-dimensional geometry.
\begin{eqnarray}\label{osmred}
ds_5^2 = - \frac{u^2}{R^2} \left( 1 - \frac{E^2R^4}{u^4} \right) d\tau^2 + \frac{R^2 u^4}{u^6 - J^2 R^6} du^2 + \frac{u^4}{R^2} \frac{E^2R^4-u^4}{J^2R^6-u^6} dx^2 + \frac{u^2}{R^2} dx_\perp^2 \ .
\end{eqnarray}
One may wonder, since the geometry described in (\ref{osmred}) approaches AdS in the UV and AdS-BH in the IR, perhaps (\ref{osmred}) can be obtained as a solution of an effective $5$-dimensional Einstein gravity with a suitable matter field
\begin{eqnarray}
S_{\rm eff} = \frac{1}{\kappa_5^2} \int d^5 x \sqrt{-G_5} \left[ R_5 - 2 \Lambda_5 \right] + S_{\rm source} \ ,
\end{eqnarray}
where $\kappa_5$ is a constant which is related to the effective $5$-dimensional Newton's constant, $G_5$ denotes the metric, $R_5$ denotes the Ricci-scalar and $\Lambda_5= - 6/R^2$ denotes the cosmological constant. Finally, $S_{\rm source}$ is the hypothetical matter field that sources the geometry (\ref{osmred}). It can be checked easily that such a putative source-term will yield a null energy condition violating energy-momentum tensor.\footnote{In this particular case, one can choose a null vector $n_\mu = (n_\tau, 1, 0, 0, 0)$ with $n_\tau^2 = \left(- g^{xx}/g^{\tau\tau} \right)$. With this one can check that $T_{\mu\nu} n^\mu n^\nu <0$.} This means that the metric in (\ref{osmred}) cannot be obtained from Einstein gravity.\footnote{This is not surprising on physical grounds: the geometry (\ref{osmred}) arises from the open string sector, there is no reason that the geometry should be sourced by a closed string sector. In fact, it is intriguing that the geometry ``knows" that it originates from open strings. Nevertheless an open string equivalence principle still holds.}

Nevertheless, an effective temperature can be read off from the metric in (\ref{osmred}). By Euclidean continuation of $\tau \to i \tau_E$, periodically compactifying the $\tau_E$ direction and demanding the absence of any conical singularity yields the following effective temperature
\begin{eqnarray} \label{temp1}
T_{\rm eff} = \sqrt{\frac{3}{2}} \, \frac{\sqrt{E}}{\pi R} \ .
\end{eqnarray}
Note that, there is no black hole in the closed string background and therefore the adjoint degrees of freedom do not experience any temperature. By virtue of the open string equivalence principle, all open string modes in the flavour sector experience an effective temperature given in (\ref{temp1}). For the sake of completeness, this effective temperature is general is given by\footnote{Here we still assume $\theta_0=0$, {\it i.e.}~the flavours are massless. For massive flavours, the general formula is more complicated and we will not write the explicit form here. It can be found in {\it e.g.}~\cite{Kim:2011qh}.}
\begin{eqnarray}
T_{\rm eff} = \frac{1}{2\pi R} \frac{\left(6 E^2 + 4 \left(\pi T R\right)^4\right)^{1/2}}{\left(E^2 + \left(\pi T R\right)^4 \right)^{1/4}} \ , \quad T = \frac{u_H}{\pi R^2} \ ,
\end{eqnarray}
where $T$ is the temperature of the close string black hole and hence of the adjoint sector in the dual field theory.

Let us now go back to (\ref{osmred}). As with a black hole geometry, the metric components are singular at $u=u_*$: the open string metric event horizon. As we approach $u \to u_*$, the radial null light cones tend to ``close up" since $d\tau/du \to \pm \infty$. This is a coordinate artifact and can be tamed with introducing the Eddington-Finkelstein coordinates. To that end, we focus on the near-horizon geometry of (\ref{osmred}) in the $\{\tau, u\}$-subspace
\begin{eqnarray} \label{osmkrus}
ds^2 = \frac{R}{24 M^2\sqrt{E}} \left[- \left(1 - \frac{2M}{\alpha}\right) d\tau'^2 + \frac{d\alpha^2}{1- \frac{2M}{\alpha}}\right] + \gamma_{xx} dx^2 + \gamma_{\perp} dx_{\perp}^2 \ , \quad \tau' = (4M)\frac{\sqrt{3E}}{L} \tau \ ,
\end{eqnarray}
where $\gamma_{xx}$ and $\gamma_{\perp}$ are numerical constants and $\alpha \to (2M)_{+}$ is the location of the horizon.

\begin{figure}[h!]
\centering
\includegraphics[scale=0.30]{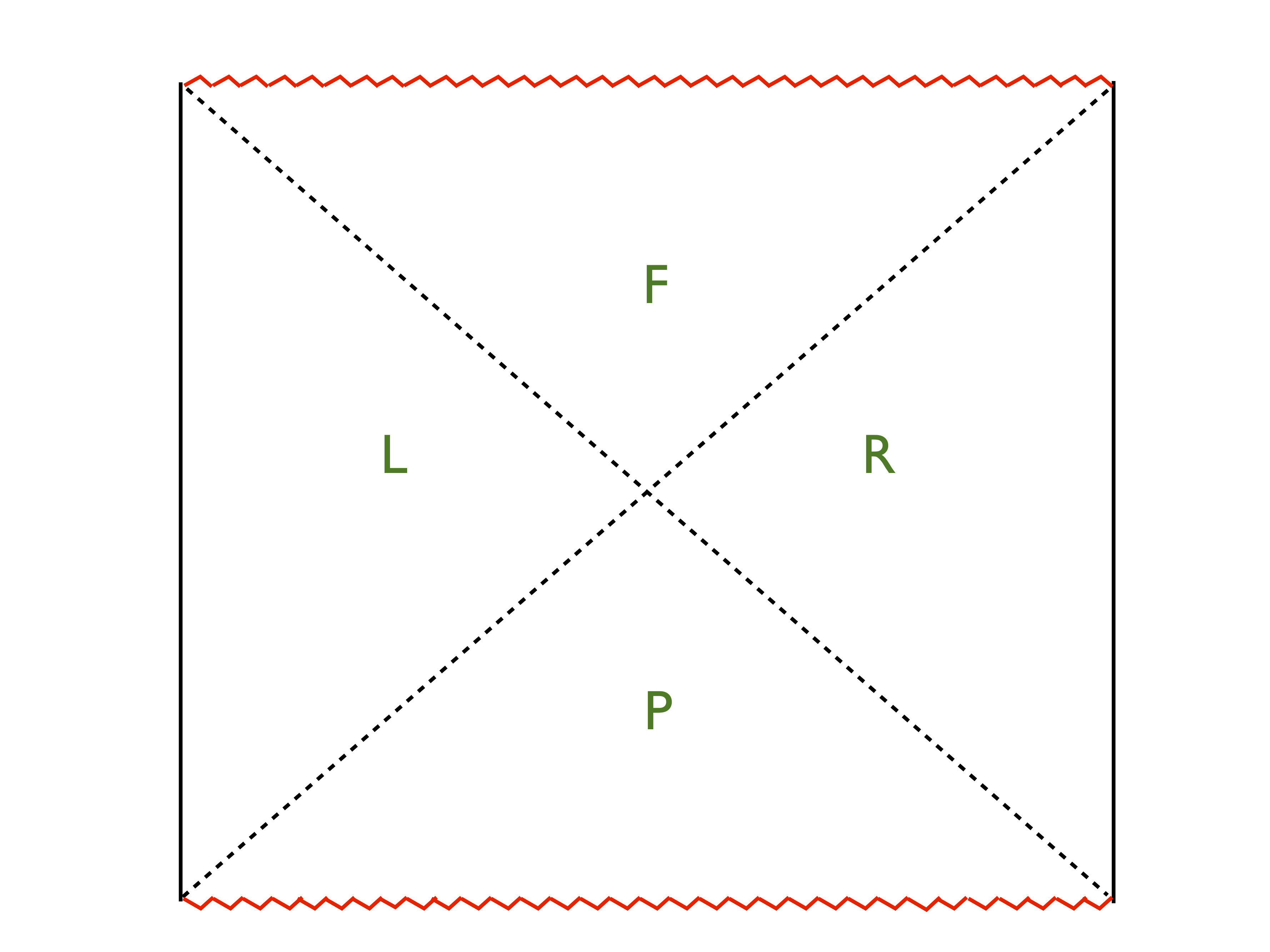}
\caption{\small A qualitative picture of the Kruskal extension of the open string metric. Note, however, that only with an AdS$_3$-background, the corresponding open string metric has a ``square" Penrose diagram, as shown above. For AdS$_{d+1}$ with $d>2$, the corresponding open string metric can also be Kruskal-extended, but the resulting Penrose diagram does not take a ``square" shape. More details will appear in \cite{KKIP}.}
\label{pen1}
\end{figure}

Following the standard lore of black holes, let us first define the Regge-Wheeler coordinate
\begin{eqnarray}
\alpha_* = \alpha + 2M \log \left| \frac{\alpha - 2M}{2M}\right| \ ,
\end{eqnarray}
and subsequently define the ingoing and outgoing null coordinates
\begin{eqnarray}
&& v = \tau' + \alpha_* \ , \quad u = \tau' - \alpha_* \ , \\
&& U = - e^{-u/4M} \ , \quad V = e^{v/4M} \ .
\end{eqnarray}
In terms of the $\{U,V\}$-patch, the metric (\ref{osmkrus}) takes the following form
\begin{eqnarray}
ds_{\rm osm}^2 = - \frac{32 M^3}{\alpha} e^{-\alpha/2M} dU dV + ds_{\perp}^2 \ .
\end{eqnarray}
where $ds_{\perp}^2$ represents all other directions which we are suppressing here. Correspondingly Kruskal-Szekres time and radial coordinates can be defined as: $t_K =  V + U$, $x_K = V - U$. Evidently, the coordinate singularity at $\alpha \to 2M_{+}$ disappears and we can extend the geometry to its Kruskal patch. The corresponding causal structure will have similar behaviour as seen in a maximally extended AdS-Schwarzschild geometry\cite{KKIP} and subsequently results in a Penrose diagram as shown in Fig.~\ref{pen1}.

The location of the event horizon $\alpha=2M$ is now given by $UV=0$, which implies either $U=0$ or $V=0$. This divides the Kruskal patch in four regions depending on the sign of $U$ and $V$ and there is a symmetry $(U,V) \to (-U,-V)$.
We will now comment on the $\tau'={\rm const}$ hypersurfaces in the Kruskal patch. To do so, we introduce the following coordinate
\begin{eqnarray}
ds_{\rm osm}^2 = - \left(\frac{1- \frac{M}{2\beta}}{1 + \frac{M}{2\beta}}\right)^2 d\tau'^2 + \left(1 + \frac{M}{2\beta}\right)^4 d\beta^2 + ds_{\perp}^2 \ , \quad \alpha = \left(1 + \frac{M}{2\beta}\right)^2 \beta \ .
\end{eqnarray}
The $\tau'={\rm const}$ hypersurfaces have topology: ${\mathbb R}_\beta \otimes {\mathbb R}_x \otimes {\mathbb R}^2$. Unlike the usual Schwarzschild case, the transverse space, described by $ds_{\perp}^2$ is insensitive to $\beta$. Nevertheless, for each value of $\alpha$, we have two values of $\beta$ which are exchanged by the symmetry: $\beta \to M^2 / 4 \beta$. The two regions, parametrized by the two roots of $\beta$, are connected by an Einstein-Rosen bridge whose ``throat" has a constant size.

\section{More examples: other brane constructions}

Motivated by the previous section, it is natural to look for a class of such examples where a universal effective temperature emerges for the probe degrees of freedom. These examples can be broadly classified in two categories: (i) Probe flavours in a conformal gauge theory and (ii) probe flavours in a non-conformal gauge theory. The examples in category (i) can be obtained by probing a D3-brane geometry by branes of various dimensions in type IIB supergravity.

Let us consider the background closed string geometry obtained by placing $N_c$ number of D3-branes on the tip of a cone, whose base is a $5$-dimensional Sasaki-Einstein manifold $\cM_5$ and taking the near-horizon limit. This gives a $10$-dimensional geometry of the form $AdS_5 \times \cM_5$. In general this will preserve $\cN=1$ supersymmetry. To begin with, let us take $\cM_5 \equiv S^5$, in which case $\cN=4$ supersymmetry is preserved. To introduce flavour degrees of freedom, we will introduce probe branes of various dimensions in the background such that these flavours are classically stable. We have schematically shown these constructions in table \ref{table2}.
\begin{table}[ht] 
\centering  
\begin{tabular}{| c | c | c | c | c | c | c | c | c | c | c | } 
\hline
Brane & 0 & 1 & 2 & 3 & 4 & 5 & 6 & 7 & 8 & 9  \\
\hline	                         
($N_c$) D3 & -- & -- & -- & -- & X & X & X & X & X & X \\
\hline                
($N_f$) D5 & -- & -- & -- & X & -- & -- & -- & X & X & X \\
\hline
($N_f$) D5 & -- & -- & X & X -- & -- & -- & -- & -- & X & X \\
\hline
\end{tabular} \caption{Flavour defects in various cases probing a D3-brane geometry with $\cM_5 \equiv S^5$.}  \label{table2}
\end{table} 
In table \ref{table2}, the notation $-$ stands for a direction along the worldvolume of the brane and $X$ represents the directions transverse to it. The directions $\{0,1,2,3\}$ represents the Minkowski directions. The rest of the directions can be parametrized in various ways, which we will do case by case. 

The Minkowski directions are shared by the $N_c$ D3-branes. For the $N_f$ probe D5-branes in the second row of table \ref{table2}, let us note that the transverse directions to the D3's is conformal to ${\mathbb R}^6 \equiv {\mathbb R}^3 \times {\mathbb R}^3$. Let us introduce spherical coordinates $\{\rho, \Omega_2\}$ and $\{\ell, \tilde{\Omega}_2\}$ to represent the ${\mathbb R}^3$'s. Now, $\{4,5,6\}$ represents $\{\rho, \Omega_2\}$ and $\{7,8,9\}$ represents $\{\ell, \tilde{\Omega}_2\}$. This configuration was studied in \cite{DeWolfe:2001pq, Filev:2009ai}. This configuration preserves $8$ real supercharges. The degrees of freedom consist of an adjoint vector multiplet and a hypermultiplet coming from the D3-branes in $(3+1)$-dimensions. The probe D5's yield a $(2+1)$-dimensional hypermultiplet in the fundamental representation of the gauge group. More explicitly, the background closed string metric can be written as
\begin{eqnarray}
&& ds^2 = ds_{AdS_5}^2  + ds_{S^5}^2 \ , \\
&& ds_{S^5}^2 = R^2 \left( d\psi^2 + \cos^2\psi d\Omega_2^2 + \sin^2 \psi d\tilde{\Omega}_2^2 \right) \ , \\
&& \rho = u \cos\psi \ , \quad \ell = u \sin\psi \ .
\end{eqnarray}
The embedding function can be parametrized by $\psi(u)$, with the condition that $P[\tilde{\Omega}_2] = 0$, where $\tilde{\Omega}_2$ denotes the metric components which yields the line element $d\tilde{\Omega}_2^2$. Exactly as in the previous section, there will be a family of solutions for the probe D5 which are characterized by the mass of the fundamental flavour. It is easy to check that we will also have a massless case --- represented by $\psi = 0$ --- which corresponds to the ``equatorial" embedding. In this case, the worldvolume geometry on the probe D5 is simply $AdS_4 \times S^2$. It can also be checked easily that we can excite the worldvolume gauge field as in the previous section and the $\psi=0$ embedding will still remain. If we focus on this embedding for simplicity, the compact directions are merely spectators.

The open string metric can be obtained to be
\begin{eqnarray}
&& ds_{\rm eff}^2 = - \frac{u^2}{R^2} \left(f - \frac{E^2R^4}{u^4}\right) d\tau^2 + \gamma_{xx} dx^2 + \gamma_{uu} du^2 + \frac{u^2}{R^2} dy^2 + \cos^2\psi d\Omega_2^2 \ , \label{osm2} \\
&& \gamma_{xx} = \frac{u^2 \cos ^4\psi \left(E^2 R^4-u^4 f \right)}{R^2 \left(j^2-u^4 f \cos ^4\psi \right)} \ , \quad   \gamma_{uu} = \frac{R^2 u^2 \cos ^4\psi \left(u^2 f \psi '^2+1\right)}{u^4 f \cos ^4 \psi - j^2} \ , \\
&& f = 1 - \frac{u_H^4}{u^4} \ .
\end{eqnarray}
Note that, evaluated on the embedding $\psi(u) =0$, the open string metric in (\ref{osm2}) yields $AdS_4 \times S^2$ only when both $u_H=0$ and $E=0$, {\it i.e.}~asymptotically. In the IR, however, we do not recover $AdS_4$-Schwarzschild geometry in the strict sense, since the power of $u$ in {\it e.g.}~$f(u)$ is reminiscent of the $AdS_5$ structure of the closed string metric. As before, requiring the positivity of $\gamma_{xx}$ the constant $j$ (which is related to the flavour conductivity in the dual field theory) can be determined in terms of $E$.\footnote{In this case one gets, $j = E R^2$, which is independent of the closed string metric temperature $T = u_H/ (\pi R^2)$.} The effective temperature is given by
\begin{eqnarray}
T_{\rm eff} = \left(T^4 + \frac{E^2}{\pi^2 R^4}\right)^{1/4} \ , \quad T = \frac{u_H}{\pi R^2} \ .
\end{eqnarray}
It is straightforward to also check that the near-horizon geometry has a very similar structure to the one obtained in (\ref{osmkrus}) and correspondingly admits a Kruskal patch. Note that the metric in (\ref{osm2}) was studied in \cite{Sonner:2012if} and it was concluded that the vector fluctuations experience an effective fluctuation-dissipation relation involving the effective temperature $T_{\rm eff}$. Here we argue that, because of the open string equivalence principle, this result holds for {\it all} fluctuations in the probe flavour sector.

We will not construct explicit examples of type (ii), but point the reader to {\it e.g.}~Sakai-Sugimoto model considered in \cite{Bergman:2008sg, Johnson:2008vna}. Such constructions can clearly be generalized for other non-conformal backgrounds.

\section{Abstracting away from the brane construction}

Motivated by the explicit examples studied in details in the previous sections, let us now ``abstract away" from the explicit $10$-dimensional construction and take a more ``phenomenological" approach. Our motivation here is to explore various features, keeping our discussion as general as possible.

\subsection{Event horizons: some generalities}

In \cite{Kundu:2013eba}, we had carried out a general analysis and obtained a general formula for $T_{\rm eff}$, the effective temperature that the steady-state system records. One intriguing question is the hierarchy of the two temperatures that we have at hand: $T$ and $T_{\rm eff}$. If $u_*$ and $u_H$ denote the open string and closed string metric horizons respectively, then it is straightforward to observe that $u_*>u_H$ in the convention where $u\to \infty$ is the boundary of the geometry. However, as pointed out in \cite{Nakamura:2013yqa} this does not imply that $T_{\rm eff}>T$. For most cases with an explicit brane construction, it is found that $T_{\rm eff} > T$.

Let us now consider a more general case, where a bunch of other gauge modes are also excited on the putative probe worldvolume. In the dual gauge theory description, these modes can correspond to introducing a non-zero chemical potential and/or a constant magnetic field in addition to the constant electric field.\footnote{See {\it e.g.}~\cite{Albash:2007bk, Erdmenger:2007bn, Mateos:2007vc, Evans:2011mu} for a detailed discussion of such physics in the D3-D7 system, and  \cite{Bergman:2008sg, Johnson:2008vna, Johnson:2009ev, Bergman:2008qv, Thompson:2008qw} for similar physics in the Sakai-Sugimoto model.} In general, we have two inequivalent combinations of the electric and the magnetic field: (i) parallel and (ii) perpendicular. For simplicity, let us take $d=3$ and $m=2$ which is sufficient to include the structures we want. We will now focus on the $\{t, x^1, x^2, x^3, u\}$-submanifold of the closed string geometry in and we will assume an $SO(3)$-symmetry in the $\{x^1,x^2,x^3\}$-plane.

Let us now consider the case when (i) $E \parallel B$. For this, the ansatz for the worldvolume gauge field is
\begin{eqnarray}
f = - E dt \wedge dx^1 - a_x' dx^1 \wedge du - B dx^2 \wedge dx^3 + a_t' dt \wedge du \ ,
\end{eqnarray}
where $a_x(u)$ contains the information about the gauge theory current and $a_t(u)$ contains the information about the chemical potential. With this ansatz we preserve an $SO(2) \subset SO(3)$ of the closed string background. The open string metric can be obtained as 
\begin{eqnarray}
 ds_{\rm eff}^2 & = & \left(\cS_{tt} - \frac{\cS_{tx}^2}{\cS_{xx}}\right) d\tau^2 + \left(\frac{\cS_{tx}}{\sqrt{\cS_{xx}}} d\tau + dX\right)^2 + \cS_{\perp} dx_{\perp}^2 , \nonumber\\
& + & \left(\frac{\cS_{tx}^2 \cS_{uu} - 2 \cS_{tu} \cS_{tx} \cS_{ux} + \cS_{tt} \cS_{ux}^2 - \cS_{tt} \cS_{uu} \cS_{xx}}{\cS_{tx}^2 - \cS_{tt} \cS_{xx}}\right) du^2  \ , \label{metabs2} \\
 dt & = & d\tau  - \frac{\cS_{tx} \cS_{ux} - \cS_{tu} \cS_{xx}}{\cS_{tx}^2 - \cS_{tt} \cS_{xx}} du \ , \\
dx^1 & = &  \frac{dX}{\sqrt{\cS_{xx}}} - \frac{\cS_{tu}\cS_{tx} - \cS_{tt} \cS_{ux}}{\cS_{tx}^2 - \cS_{tt} \cS_{xx}} du \ , 
\end{eqnarray}
with
\begin{eqnarray}
&& \cS_{\perp} = G_{xx} + \frac{B^2}{G_{xx}}   \ , \quad dx_{\perp}^2 = \left(dx^2\right)^2 + \left(dx^3\right)^2 \ ,
\end{eqnarray}
and 
\begin{eqnarray}
&& \cS_{tt} = G_{tt} + \frac{E^2}{G_{xx}} + \frac{a_t'^2}{G_{uu}} \ , \quad \cS_{tx} = \frac{a_t' a_x'}{G_{uu}} \ , \quad \cS_{tu} = \frac{E a_x'}{G_{xx}} \ , \\
&& \cS_{xx} = G_{xx} + \frac{E^2}{G_{tt}} + \frac{a_x'^2}{G_{uu}} \ , \quad  \cS_{uu} = G_{uu} + \frac{a_t'^2}{G_{tt}} + \frac{a_x'^2}{G_{uu}} \ .
\end{eqnarray}
The location of the open string metric event-horizon is given by
\begin{eqnarray}
 \cS_{tx}^2 - \cS_{tt} \cS_{xx} = 0  \quad & \implies& \quad \left. \tilde{d}^2 e^{2\phi} G_{tt} + G_{xx} \left( B^2 G_{tt} + G_{tt} G_{xx}^2 + e^{2\phi} j^2 \right) \right|_{u_*} = 0 \nonumber\\
& \implies & \quad  \left.G_{tt}G_{xx} + E^2 \right|_{u_*} = 0 \ .
\end{eqnarray}
Here the constant $j$ and $d$ are defined as
\begin{eqnarray}
j = \frac{\partial \cL}{\partial a_x'} \ , \quad \tilde{d} = \frac{\partial \cL}{\partial a_t'} \ ,
\end{eqnarray}
where $\cL$ represents the DBI Lagrangian of the probe. The constant $j$ is related to the boundary current and $\tilde{d}$ is related to the chemical potential of the system. The constant $j$ can be fixed by requiring that the coordinate $X$ does not change sign.

For completeness, let us now comment on the case when (ii) $E \perp B$. In this case, there will be an additional current due to the Hall effect, see {\it e.g.}~\cite{O'Bannon:2007in}. The worldvolume gauge field ansatz will take the form
\begin{eqnarray}
f = - E dt \wedge dx^1 + B dx^1 \wedge dx^2 - a_x' dx^1 \wedge du - a_y' dx^2 \wedge du + a_t' dt \wedge du  \ .
\end{eqnarray}
The open string metric can be obtained as before and there is an event-horizon which can subsequently be obtained. We refrain from providing the explicit expressions since they are complicated looking and not particularly illuminating.

\subsection{Ergoplane: chemical potential}

Let us now focus on a special case, where the open string metric acquires an ``ergoplane" and an event horizon. This was explicitly demonstrated in the D3-D7 system in \cite{Kim:2011qh}. Here we will argue that a similar structure persists more generally. To this end, we consider only an electric field along with the chemical potential. By setting $B=0$ in (\ref{metabs2}), we get 
\begin{eqnarray}
ds_{\rm eff}^2 & = & \left(G_{tt}G_{xx} +E^2\right) \left[ \frac{dT^2}{G_{xx}} + \frac{dX^2}{G_{xx}} \right] + \frac{1}{G_{uu}} \left( a_t' dT + a_x' dX \right)^2 + \sum_{i=2}^m G_{xx} dx^i dx^i \nonumber\\
&+& \left(G_{uu} + \frac{a_t'^2 G_{tt} + a_x'^2 G_{xx}}{G_{tt} G_{xx} + E^2}\right) du^2 \ , \\
dT & = & dt + \frac{E a_x'}{G_{tt} G_{xx} + E^2} du \ , \quad dX= dx - \frac{E a_t'}{G_{tt} G_{xx} + E^2} du \ .
\end{eqnarray}
It is clear that the event horizon of this metric obtained from $(G_{tt} G_{xx} + E^2) = 0$, but there is an ergoplane in this geometry which is obtained by
\begin{eqnarray}
 \frac{G_{tt}G_{xx} + E^2}{G_{xx}} + \frac{a_t'^2}{G_{uu}} = 0  \quad & {\implies}& \quad \left(G_{tt}G_{xx} + E^2\right) = 0  \ , \nonumber\\
& {\rm or} & \quad \left(G_{tt} G_{xx}^m + e^{2\phi} j^2 \right) = 0 \ .
\end{eqnarray}
The first condition gives the open string metric event horizon, but the second condition yields a root, denoted by $u_{\rm erg}$ henceforth, which is different from $u_*$. We also note that $j^2$ is determined from the algebraic condition: $\left. \tilde{d}^2 e^{2\phi} G_{tt} + \left(G_{tt} G_{xx}^m + e^{2\phi} j^2 \right) G_{xx} \right|_{u_*}= 0$.

If the background represents an AdS$_{d+2}$ geometry, in which the dilaton $\phi=0$, then it can readily seen that 
\begin{eqnarray}
u_{\rm erg}^2 = E L^2 \left( 1+ \frac{\tilde{d}^2}{E^d}\right)^{1/d} > u_*^2 = E L^2 \ , 
\end{eqnarray}
where we have used the conditions to determine $u_*$, $j$ and $u_{\rm erg}$. Evidently, the event horizon and the ergoplane merge in the limit $(\tilde{d}^2/E^d) \to 0$ and/or $(1/d) \to 0$.\footnote{This inequality $u_{\rm erg}>u_*$ can be shown to hold for the AdS$_{d+2}$-Schwarzschild geometry as well, and $u_{\rm erg} \to u_*$ in the limit $\tilde{d} \to 0$ and/or $d \to \infty$.}

\subsection{Exactly solvable toy examples}

We will now discuss a few cases where {\it e.g.}~the gauge field fluctuations can be solved analytically. Towards that end, let us take a generic Lifshitz background of the following form:
\begin{eqnarray}
ds^2 = - \frac{dt^2}{v^{2z}} + \frac{1}{v^2} d\vec{x}^2 + \frac{dv^2}{v^2} \ ,
\end{eqnarray}
where $\vec{x}$ is a $2$-dimensional vector and $v$ is the radial coordinate ($v \to 0$ corresponds to boundary and $v\to\infty$ corresponds to the deep-IR). In this background we will consider a probe of the same bulk space-time dimensions whose action is given by the DBI action.\footnote{For now, we will assume that the full $10$ or $11$-dimensional background (if it exists) allows for such a probe brane which has a constant profile along the compact directions. Thus the effective lower dimensional problem that we have to solve is done by simply considering the ``reduced" DBI action of the appropriate dimensions.}

Moreover, to introduce the ``steady-state" physics that we are interested in, we introduce a gauge field on the world volume of the probe given by
\begin{eqnarray}
a_x = - E t + h(v) \ . 
\end{eqnarray}
The equation of motion for the function $h(v)$ is given by
\begin{eqnarray}
\frac{v^{1-z} h'}{\sqrt{1-e^2 v^{2+2z} +v^4 h'^2 }} = j \ ,
\end{eqnarray}
where $e = (2\pi\alpha') E$ and $j$ is the constant related to the boundary current. The on-shell Lagrangian using this equation of motion is obtained to be
\begin{eqnarray}
\cL_{\rm os} = \frac{1}{v^{3+z}} \sqrt{\frac{e^2 v^{2+2z} -1 } {j^2 v^{2+2z} -1}} \ .
\end{eqnarray}
From the above expression, imposing the reality condition for the on-shell action, we get $j = e$ which simply yields $\cL_{\rm os} = 1/ v^{3+z}$.

Now, we can consider the gauge field fluctuations living on the world volume of the probe: $a = a_0 + \delta a$, where $a_0$ denotes the classical profile of the gauge field. The fluctuation modes will sense the OSM. This fact is reflected in the action for the quadratic fluctuation for the gauge fields which is given by
\begin{eqnarray}
S_{\rm gauge} = - \frac{1}{4} \int dt d^dx \sqrt{- {\rm det}(G+f)} \cS^{aa'} \cS^{bb'} \delta f_{ab} \delta f_{a'b'} \ ,
\end{eqnarray}
where $\sqrt{-{\rm det} (G+f)} = \cL_{\rm os}$ and $\delta f = d \delta a$ is the field strength of the gauge field fluctuations. This action results in the following equation of motion
\begin{eqnarray}
\partial_a \left[\sqrt{-{\rm det} (G+f)} \cS^{aa'} \cS^{bb'} \delta f_{a'b'} \right] = 0 \ .
\end{eqnarray}
In the $\{t,v\}$-plane the OSM is non-diagonal and we can diagonalize it by introducing an ingoing Eddington-Finkelstein type coordinate
\begin{eqnarray}
\tau = t + s(v) \ , \quad \frac{ds}{dv} = -\frac{e^2 v^{1+ 3z}} {1- e^2 v^{2z+2}}
\end{eqnarray}
where $\tau$ is the new time coordinate. This gives
\begin{eqnarray}
ds^2 = - \frac{g(v)}{v^{2z}} d\tau^2 + \frac{1}{g(v)} \frac{dv^2}{v^2} \ , \quad g(v) = 1 - e^2 v^{2z+2} \ .
\end{eqnarray}

Now, the vector fluctuations that we consider are of the following form
\begin{eqnarray}
\delta a_\tau = \delta a_\tau (v) e^{-i\omega \tau} \ , \quad \delta a_i = \delta a_i (v) e^{- i \omega \tau} \ . 
\end{eqnarray}
It is straightforward to check that the temporal and the spatial fluctuations decouple and hence we study only the spatial fluctuations here.\footnote{Interestingly the vector fluctuations along both the $x$-direction and the $y$-direction obey the same equation.} The general equation for the spatial vector fluctuations take the following form
\begin{eqnarray}
\partial_v \left[ \frac{v}{v^z} g(v) \partial_v \delta a_x \right] + \frac{\omega^2}{g(v)} \frac{v^z}{v} \delta a_x = 0 \ .
\end{eqnarray}
Interestingly, we can solve the above equation analytically for $z = 1/2, \, 1, \, 3/2, \, 2$. It will be interesting to see what happens to the fluctuation-dissipation relation in all these cases.

Some of the explicit solutions take the following form:
\begin{eqnarray}
 z & = &  \frac{1}{2} \ , \nonumber\\
 \delta a_x(v) & = &  c_x \, \left( \frac{1 - \sqrt{v}}{1+ \sqrt{v}} \right)^{- \frac{i \omega}{3}} \left( \frac{1 - \sqrt{v} + v }{1+ \sqrt{v} + v } \right)^{- \frac{i \omega}{6}} \, {\rm exp} \left( \frac{i\omega}{\sqrt{3}} \arctan\left[  \frac{\sqrt{3} v}{1- v}\right] \right) \ ,  \label{z12} \\
 z & = &  1 \ ,  \nonumber\\
\delta a_x(v) &  = & c_x \,  \left( \frac{1 - v}{1+v} \right)^{- \frac{i \omega}{4}}   \, {\rm exp} \left( \frac{i\omega}{2} \arctan (v) \right) \ . \label{z1} \\
z & = & 2 \ , \nonumber \\
\delta a_x(v) & = & c_x \, \left( 1- v^2 \right)^{\frac{ - i \omega}{6}} \left( 1 + v^2 + v^ 4  \right)^{\frac{i \omega}{12}}  \, {\rm exp} \left( \frac{i\omega}{2\sqrt{3}}  \arctan \left( - \frac{\sqrt{3}}{1 + v^2} \right) \right) \ . \label{z2}
\end{eqnarray}
Here $c_x$ is a constant. Using (\ref{z12})-(\ref{z2}), the asymptotic form of the gauge field can be easily determined
\begin{eqnarray}
\delta a_x ( v) = c_x \left( 1 +  i \omega \Lambda^{(1)} v^{z} + \ldots \right)  \ ,
\end{eqnarray}
where $\Lambda^{(1)}$ is a numerical constant. Using this asymptotic expansion, the retarded current-current correlator is easily obtained to be
\begin{eqnarray}
G_{\rm R}^{ij} \equiv \lim_{\epsilon \to 0} \frac{\partial^2 S_{\rm gauge}}{ \partial \left(\delta a_i (\epsilon) \right) \partial \left( \delta a_j(\epsilon) \right)} = - \left( i z \omega \right)\Lambda^{(1)}  \delta^{ij} \ ,
\end{eqnarray}
where $\{i,j\}$ takes values $\{x,y\}$.

Now, using the Schwinger-Keldysh method in the context of gauge-gravity duality, as pioneered in \cite{Herzog:2002pc}, we can easily arrive at
\begin{eqnarray}
G_{\rm sym}^{ij} = - \left( 1 + 2 n_{\rm eff} \right) {\rm Im} G_{\rm R}^{ij} \ , \quad n_{\rm eff} = \frac{1}{{\rm exp}\left( \frac{\omega}{T_{\rm eff}} \right) -1 } \ ,
\end{eqnarray}
where $G_{\rm sym}^{ij}$ denotes the symmetric Schwinger-Keldysh correlator. Clearly, we observe that the fluctuations indeed measure a temperature $T_{\rm eff}$.\footnote{Recall that, even though the analysis in \cite{Herzog:2002pc} was done for a scalar field, the corresponding analysis for a gauge field is similar. Ultimately, one needs to explore the analyticity of positive frequency modes as the event horizon is crossed, be it the background event horizon or the open string metric event horizon. Therefore we get an analogous result.} In this case, we get
\begin{eqnarray}
T_{\rm eff} = \frac{z+1}{\pi} e^{z/ (z+1)}  \ .
\end{eqnarray}
%

\subsection{The probe limit and its validity}

It is clear from the above discussion that, in order for us to have the above structure, we need to treat an additional set of D-branes in a background which is obtained by solving supergravity. Let us imagine that the following geometric data represents a particular solution of $10$-dimensional supergravity in string frame
\begin{eqnarray}
ds^2 & = & G_{tt}(u) dt^2 + G_{xx}(u) d\vec{x}_p^2 + G_{uu}(u) du^2 + G_{c}(u) d\Omega_{8-p}^2 \ , \label{metgenp} \\
\phi & = & \phi(u) \ . \label{dilgenp}
\end{eqnarray}
Here all the symbols have their usual meaning, specially $d\Omega_{8-p}^2$ stands for an $(8-p)$-dimensional compact manifold. Quite clearly, the putative geometry in (\ref{metgenp}) is dual to a $(p+1)$-dimensional gauge theory, which possess $SO(p)$ rotational invariance and is also homogeneous in the spatial coordinates. Furthermore, by assumption, there is a global symmetry which is the isometry group of the manifold whose line element is represented by $d\Omega_{8-p}^2$. Evidently, (\ref{metgenp})-(\ref{dilgenp}) may not characterize the full geometric data that can further contain various non-vanishing form-fields. However, for our purposes, these form fields will not play any role as we will argue momentarily.

Before going further, let us note that the background in (\ref{metgenp}) source an Einstein tensor
\begin{eqnarray}
E_{\mu\nu} = R_{\mu\nu} - \frac{1}{2} G_{\mu\nu} R \ .
\end{eqnarray}

Let us now imagine that we place a D$q$-brane in the above background in the {\it probe limit}. Let us imagine that this probe brane extends along ${\mathbb R}^m \subset {\mathbb R}^p$, the radial direction $u$ and wraps an $\cM_{q-m-1} \subset \cM_{8-p}$. The last assumption imposes: $ \left( p + q \right) \le \left( 9 + m \right) $.

Now, to describe the steady-state, we excite a gauge field along one of the spatial directions as before
\begin{eqnarray}
A_{y} = - E t + a_y(u) \ .
\end{eqnarray}
On the worldvolume of the probe brane this breaks the $SO(m) \to SO(m-1)$. The resulting action is given by
\begin{eqnarray}
S_{\rm probe} = - T_{{\rm D}q} \int e^{-\phi} G_{xx}^{(m-1)/2} G_{c}^{(q-m-1)/2} \left[ G_{uu} \left( |G_{tt}| G_{yy} -E^2 \right) + |G_{tt}| a_y'^2 \right]^{1/2} \ .
\end{eqnarray}
We have further assumed that there is no Wess-Zumino term for the above configuration.

The gauge field equation now gives
\begin{eqnarray}
a_y' = J e^{\phi} \left(\frac{G_{uu} G_{xx} G_c^{m+1}}{|G_{tt}|} \right)^{1/2} \sqrt{\frac{|G_{tt}| G_{yy} - E^2}{G_c^q |G_{tt}| G_{xx}^m - J^2 e^{2\phi} G_{c}^{m+1} G_{xx}}} \ ,
\end{eqnarray}
where $J$ is a constant.

Now, the {\it probe limit} condition is satisfied if
\begin{eqnarray}
|E_{\mu\nu}| \gg |T_{\mu\nu}| \ .
\end{eqnarray}
This condition is non-trivial, since it involves the radial coordinate $u$.

To flesh out the idea, let us now consider a set of examples. Imagine we start from the geometry which is obtained by taking a near-horizon limit of $N_c$ coincident D$p$-branes. This, in the string frame, is given by\cite{Itzhaki:1998dd}
\begin{eqnarray}
ds^2 & = & \left(\frac{u}{L}\right)^{\frac{7-p}{2}} \left ( - dt^2 + d\vec{x}_p^2\right ) + \left(\frac{L}{u}\right)^{\frac{7-p}{2}} \left( du^2 +  u^2 d\Omega_{8-p}^2 \right) \ , \\
e^\phi & = & \left(\frac{u}{L}\right)^{\frac{(p-3)(7-p)}{4}} \ , \\
F_{8-p} & = & \left( 7 - p\right) L^{7-p} \omega_{8-p} \ .
\end{eqnarray}
Here $d\Omega_{8-p}$ denotes the line element of an $(8-p)$-sphere and $\omega_{8-p}$ denotes the corresponding volume form. This background preserves $16$-supercharges. It is straightforward to check that for the above geometry, the Einstein tensor behaves as:
\begin{eqnarray}
E_{tt} \sim \frac{u^{5-p}}{L^{7-p}} \sim E_{xx} \ , \quad E_{uu} \sim \frac{1}{u^2} \ , \quad E_{\alpha\beta} \sim u^{0} \, \eta_{\alpha\beta}  \ . \label{testbem}
\end{eqnarray}

Now, let us imagine introducing an $N_f$-number of probe D$(p+4)$-branes that extend along the $\{t, \vec{x}_p\}$-directions and wraps three-cycle $\cX_3 \subset S^{8-p}$. This, in general, preserves $\cN=1$ supersymmetry. Furthermore, we excite a gauge field along {\it e.g.}~$x^1$-direction that breaks $SO(p) \to SO(p-1)$:
\begin{eqnarray}
A_{x^1} = - E t + a_1 (u) \ .
\end{eqnarray}
It is straightforward now to check, using the formulae (\ref{Tprobe1})-(\ref{Tprobe5}), that the probe energy-momentum tensor yields
\begin{eqnarray}
T_{tt} \to \frac{u^2}{L^3} \sim T_{xx} \sim T_{x^1 x^1} \ , \quad T_{uu} \to \frac{1}{u^{5-p}} \ , \quad T_{\alpha\beta} \to \frac{1}{u^{3-p}} \, \eta_{\alpha\beta} \quad {\rm as} \quad u\to \infty \ , \label{testpemUV}
\end{eqnarray}
and
\begin{eqnarray}
T_{tt} & \to & \left( E J \right) u^{(p-9)/2} \ , \quad T_{x^1x^1} \to \left( \frac{J}{E} \right) u^{(5-p)/2} \ , \quad T_{xx} \to \left( \frac{E}{J} \right) u^{(p+3)/2} \ , \\
T_{uu} & \to & \left( E J \right) u^{(3p-23)/2} \ , \quad T_{\alpha\beta } \to \left( \frac{E}{J} \right) u^{(3p-7)/2} \quad {\rm as} \quad u \to 0 \ .  \label{testpemIR}
\end{eqnarray}
Comparing (\ref{testbem}) with (\ref{testpemUV}), we observe that near the UV, the probe limit is perfectly fine, except the case with $p=4$. Similarly, comparing (\ref{testbem}) with (\ref{testpemIR}), we observe that the deep infra-red of the background will be modified by the probe sector. Further note that, $T_{tt} \sim (E \cdot J)$, which amounts to the Ohmic dissipation term from the perspective of the boundary field theory.

\subsection{A few words on the effective thermodynamics}

We have argued in the previous sub-section that, the fluctuation modes will sense an effective temperature $T_{\rm eff}$, which is an intensive variable. The thermodynamically conjugate variable is entropy, which is extensive and dynamical. As argued in \cite{Karch:2008uy}, only thermodynamic quantity that can be reliably calculated in the probe limit is the free energy, which is simply given by the on-shell action.

The basic idea of gauge-gravity duality equates the {\it bulk gravity path integral} with the dual field theory path integral, {\it i.e.}
\begin{eqnarray}
&& \cZ_{\rm bulk} = \cZ_{\rm QFT} \ , \quad \cZ_{\rm bulk} = e^{i S_{\rm bulk}} \ , \quad S_{\rm bulk} = S_{\rm grav} + S_{\rm probe} \ , \nonumber \\
&& {\rm subsequently} \quad \cZ_{\rm QFT} = e^{i S_{\rm QFT}} \ , \quad  S_{\rm QFT} = S_{\rm adj} + S_{\rm fund} \ .
\end{eqnarray}
Thus, in the probe limit {\it i.e.}~$N_f \ll N_c$, the probe sector path integral will give the path integral for the fundamental sector in the dual field theory. In Euclidean signature, this statement becomes an equality of the partition functions of the two sides. However, as \cite{Karch:2008uy} has emphasized already, this is useful as long as we calculate the free energy. 

Now, a proposal to obtain the thermodynamic free energy was put forward in \cite{Alam:2012fw}, which was subsequently generalized in \cite{Kundu:2013eba}. According to this proposal, the Helmholtz free energy is given by
\begin{eqnarray}
\cF_{\rm H} & = & \left. T_{\rm eff} \, S_{\rm DBI}^{(\rm E)} \right |_{\rm on-shell} \nonumber \\
& \sim & \int_0^{u_*} du d^p\xi \left( \cL_{\rm on-shell} - j a_x' \right) \ .
\end{eqnarray}
Here $u_*$ is the osm event-horizon, $S_{\rm DBI}^{(\rm E)}$ is the Euclidean DBI action. This is the only extensive quantity that we can define.

On the other hand, given the existence of an osm event-horizon, and the open string data $\{\cG_s, \cS \}$, where $\cG_s$ is the open string coupling\cite{Seiberg:1999vs}, there is an area associated and therefore a natural notion of {\it some entropic quantity}, given by
\begin{eqnarray}
s \sim \left. \frac{1}{\cG_s} {\rm Area} \left( \cS \right) \right |_{\tau= {\rm const}, \, u = u_*} \ .
\end{eqnarray}
However, it is not clear to us at present what this quantity physically corresponds to. One possibility could be this measures the entanglement entropy for the pair creation process that drives the steady-state current (similar to \cite{Jensen:2013ora}), but we leave this for future work.

\subsection{A dynamical horizon}

So far, we have observed stationary space-time described by the open string metric. Here we will review an example where the open string metric yields a dynamical geometry following \cite{Karch:2010kt}. Let us imagine for some given closed string metric, the induced metric on the probe worldvolume takes the effective form
\begin{eqnarray} \label{EFmet}
ds^2 = - u^2 f(u) dV^2 + 2 dV du + u^2 dx_d^2 \ ,
\end{eqnarray}
where we have introduced the Eddington-Finkelstein ingoing coordinate
\begin{eqnarray}
dt = dV + h(u) du \ , \quad {\rm with} \quad h(u) = - \frac{1}{u^2 f(u)} \ .
\end{eqnarray}
The metric takes the usual black hole metric form in the $\{t, u\}$-patch. The corresponding DBI action takes the following schematic form
\begin{eqnarray}
S_{\rm DBI} = - \tau \int d^{d+2} \xi e^{-\phi} G_{xx}^{d/2} \, X \ , \quad X = \left[ 1+ G_{xx}^{-1} \left( 2 f_{xu} f_{xV} - f_{xu}^2 G_{tt}\right)\right]^{1/2} \ ,
\end{eqnarray}
where we have considered a generic field strength of the world volume $f = f_{xu}(u) dx \wedge du + f_{xV} (u) dx \wedge dV$, which is capable of capturing the physics we want. Subsequently, we will get the following equation of motion
\begin{eqnarray}
&& \partial_u \left[\frac{e^{-\phi} G_{xx}^{(d-2)/2} \left( f_{xV} - f_{xu} G_{tt}\right)}{X}\right] + \partial_V \left[\frac{e^{-\phi} G_{xx}^{(d-2)/2} f_{xu}}{X}\right] = 0 \ , \\
&& \partial_u f_{xV} + \partial_V f_{ux} = 0 \ .
\end{eqnarray}
The second condition above arises from the Bianchi identity.

A simple solution of the above equations is given by\cite{Karch:2010kt}
\begin{eqnarray}
f_{xu} = 0 \ , \quad f_{xV} = E(V) \ , \quad {\rm provided} \quad \partial_u \left[\frac{e^{-\phi} G_{xx}^{(d-2)/2}}{X}\right] = 0 \ . \label{dynhori}
\end{eqnarray}
The last condition in (\ref{dynhori}) is trivially satisfied if (\ref{EFmet}) is an AdS$_4$ metric with a vanishing dilaton. The corresponding open string metric now takes the form
\begin{eqnarray}
ds_{\rm osm}^2 = \left(G_{tt} + \frac{E(V)^2}{G_{xx}}\right) dV^2 + 2 dV du + u^2 dx_2^2 \ , \label{osmdyn}
\end{eqnarray}
which has an AdS-Vaidya form. Choosing $E(V)$ to be a smooth function interpolating from zero to a non-vanishing finite value, we can simulate the formation of an event horizon in the open string metric. In the dual field theory this process of horizon formation encodes turning on a time-dependent boundary current $j(t) \sim E(t)$. At each $t = {\rm const}$ slice, the geometry in (\ref{osmdyn}) has an apparent horizon located at $\left(G_{tt} + E(V)^2/G_{xx}\right) = 0$.

\section{Schrodinger symmetric background}

We will now briefly discuss one special case which does not fall under the metric class that we have discussed so far. This special case is the metric with Schr\"{o}dinger isometry. This geometry can be obtained by a non-relativistic deformation of the relativistic AdS-background\cite{Son:2008ye, Balasubramanian:2008dm} in {\it e.g.}~type IIB supergravity
\begin{eqnarray}
&& ds^2 = - \frac{\Omega}{z^{2n}} \left(dX^+\right)^2 + \frac{1}{z^2} \left(- 2 dX^+ dX^- + dx_1^2 + dx_2^2 + dz^2 \right) + ds_{X^5}^2 \ , \label{schrmet} \\
&& B_2 = \frac{1}{z^n} \cA \wedge dX^+ \ , \quad  F_5 = \left(1 + \star \right) {\rm Vol}_{X^5} \ ,
\end{eqnarray}
where $z$ is the radial direction, $X^5$ is an Einstein manifold and $\cA$ is a vector eigenfunction of the Laplacian on $X^5$ and $n$ is determined by the corresponding eigenvalue. The metric function $\Omega$ obeys an inhomogeneous scalar Laplace equation on the internal manifold. When $X^5$ is also a Sasaki manifold, it is possible to preserve at least two real supercharges depending on the choice of $\cA$\cite{Bobev:2009mw}. It is easy to see that (\ref{schrmet}) arises from a deformation of an underlying AdS-metric. Turning off $\Omega$ and $\cA$ and identifying 
\begin{eqnarray}
X^+ = t + y \ , \quad X^- = \frac{1}{2} \left(t - y \right) \ ,
\end{eqnarray}
we recover an AdS$_5$ geometry, where $X^+$ and $X^-$ are identified with the null coordinates.

The metric (\ref{schrmet}) is invariant under the Schrodinger group, denoted by Schr$(d)$ in $d$-spatial dimensions. This group is an analogue of the relativistic conformal group and has the following generators: temporal translation, spatial translations, rotations, Galilean boosts, dilatation and a special conformal transformation when $n=2$. The index $n$, which encodes the information of how spatial and the time directions in the dual field theory scale, is known as the dynamical exponent. Note that, the dual field theory lives in $2$-dimensions less: the time coordinate is identified with $X^+$ and $x_1$ and $x_2$ are the spatial directions.

For concreteness we will now focus on the case where $X^5\equiv S^5$ and $n=2$\cite{Herzog:2008wg,Maldacena:2008wh,Adams:2008wt}, which can be written as an $U(1)$ fibration over a K\"{a}hler-Einstein base: $\mathbb{CP}^2$:
\begin{eqnarray}
ds_{S^5}^2 = \left(d\chi + \cA\right)^2 + ds_{{\mathbb CP}^2} \ , \quad ds_{{\mathbb CP}^2} = d\theta^2 + \cos^2 \theta \left(\sigma_1^2 + \sigma_2^2 + \sin^2 \theta \sigma_3^2 \right) \ , 
\end{eqnarray}
where $\sigma_i$, $i=1,2,3$ are the $SU(2)$ left-invariant one-forms (following notations of \cite{Ammon:2010eq})
\begin{eqnarray}
&& \sigma_1 = \frac{1}{2} \left( \cos \alpha_2 d\alpha_1 + \sin\alpha_1 \sin \alpha_2 d\alpha_3 \right) \ , \quad \sigma_2 = \frac{1}{2} \left( \sin \alpha_2 d\alpha_1 - \sin\alpha_1 \cos \alpha_2 d\alpha_3 \right) \ , \nonumber\\
&& \sigma_3 = \frac{1}{2} \left( d\alpha_2 + \cos\alpha_1 d\alpha_3 \right) \ .
\end{eqnarray}
Also, the vector field is given by $\cA = \cos^2 \theta \sigma_3$. In this case, the background above can be obtained from an AdS$_5 \times S^5$ geometry by performing the null Melvin twist (nMT) procedure. Thus, an {\it intuitive} way to embed the D7-branes in the Schr\"{o}dinger geometry will be to consider an embedding analogous to section 2\cite{Ammon:2010eq}. This means that the D7-brane wraps the following directions: $\{X^+, X^-, x_1, x_2, z, \alpha_1, \alpha_2, \alpha_3\}$ and $\theta(z)$ and $\chi(z)$ are the transverse scalars. Without any loss of generality, we can set $\chi(z) = 0$, since it has an $U(1)$-symmetry. Now, it can be easily checked that the DBI action yields exactly the same result as in the purely AdS$_5\times S^5$ case.\footnote{This is no accident. The fact that the Schr\"{o}dinger background is obtained by the nMT procedure implies that the DBI corresponding to the D7-brane embedded above remains invariant under the nMT transformations. It can also be checked that the corresponding open string metric on the D7-brane is AdS$_5 \times S^3$ for the $\theta=0$ embedding. On the other hand, for a generic dynamical exponent $n\not =2$, the open string metric is a null deformation of the AdS-metric: $ds_{\rm osm}^2 = - z^{-2n} \left(\Omega - ||\cA||^2\right) \left(dX^+\right)^2 + ds_{\rm AdS}^2 + \ldots $.} Hence, the simplest embedding --- which corresponds to introducing massless flavors --- is again given by: $\theta(z) =0$.

To excite the gauge field analogous to (\ref{gaugeansatz}), we consider\cite{Ammon:2010eq}
\begin{eqnarray}
A_{x_1} = - \frac{1}{2} E X^+ - E X^- + a_x(z) \ .
\end{eqnarray}
Following the similar chain of arguments, we will now have the open string event horizon and the boundary flavor current 
\begin{eqnarray}
z_* ^2 = \frac{1}{E} \ , \quad j = E^{3/2} \ , 
\end{eqnarray}
and finally the open string metric 
\begin{eqnarray} \label{schrosm}
ds_{\rm osm}^2 & = & - \frac{1 - E^2 z^4}{z^2} d\tau^2 + \frac{1}{z^2} \left(dx_2^2 + dy^2\right) + \frac{z^6 E^4 + z^4 E^3 + z^2E^2 + z^2 E + 1}{z^6 E^2 + z^4 E + z^2} dx_1^2 \nonumber\\
&+& \frac{dz^2}{z^2 \left( 1 - E^3 z^6 \right)} + ds_{S^3}^2 - 2 E dx_1 \sigma_3 \ ,
\end{eqnarray}
where we have redefined 
\begin{eqnarray}
d\tau = d t - h(z) dz \ , \quad h(z) = \frac{E^{5/2} z^5}{ 1- E^3 z^6} \sqrt{\frac{z^4 E^2 + z^2 E + 1}{z^2 E + 1}} \ .
\end{eqnarray}
As before, we observe that the open string metric acquires an event horizon at $z_* = 1/\sqrt{E}$ in a similar manner as before. As before, the fluctuation modes will be governed by an open string equivalence principle and an effective thermodynamic description will prevail.

\section{Further generalization: non-Abelian DBI}

Let us begin with the non-Abelian version of the Dirac-Born-Infeld action\cite{Myers:1999ps}. In much of what we present below, we will closely follow the notations and conventions of \cite{Erdmenger:2008yj}. This reads
\begin{eqnarray}
S_{\rm DBI} = - N_f T_{{\rm D}7}{\rm STr} \int d^8 \xi \sqrt{{\rm det} Q} \left [ {\rm det} \left( P_{ab} \left[ E_{\mu\nu} + E_{\mu i} \left( Q^{-1} - \delta \right)^{ij} E_{j \nu} \right] + 2 \pi\alpha' F_{ab} \right) \right]^{1/2} \ , \label{nonAbe}
\end{eqnarray}
where
\begin{eqnarray}
Q_i^j & = & \delta_i^j + i \left( 2 \pi \alpha' \right) \left[ \Phi^i, \Phi^j \right] E_{ij} \ , \\
E_{\mu\nu} & = & G_{\mu\nu} + B_{\mu\nu} \ .
\end{eqnarray}
Furthermore, $\mu, \nu = 0, \ldots , 9$; $a,b = 0, \ldots, 7 $ and $i,j = 8,9$, which are all space-time indices. Also, here ${\rm STr}$ corresponds to ``symmetrized trace" and $\Phi^i$ denotes the profile function of the brane. Evidently, the flavour index is suppressed in the above expressions.

To simplify the situation, we imagine placing $N_f$ probe D$7$-branes in an AdS-background. Thus, we can make use of the rotational symmetry along the $\{8,9\}$-directions to set $\Phi^9 = 0$. Thus we get $Q_i^j = \delta_i^j$. Furthermore, here $E_{\mu\nu} = G_{\mu\nu}$ since the NS-NS anti-symmetric form $B_{\mu\nu} = 0$. Thus, to evaluate (\ref{nonAbe}), we need to compute only the pull-back of the background metric, which is given by
\begin{eqnarray}
P_{ab} \left[ G \right] = G_{ab} + \left( 2 \pi \alpha' \right)^2 G_{ij} \left(\partial_a \Phi^i \partial_b \Phi^j + i \partial_{(a} \Phi^i \left[ A_{b)}, \Phi^j \right]  - \left[ A_a, \Phi^i \right]  \left[ A_b, \Phi^j \right] \right) \ . \label{pbnonAbe}
\end{eqnarray}

Let us now focus on the case of $N_f=2$. In this case, the corresponding group $U(2)$ is generated by a set of $4$ ${\mathbb C}$-valued $2 \times 2$ matrices:
\begin{eqnarray}
\sigma^0 = \left( \begin{array}{cc}
1 & 0 \\
0 & 1 \end{array} \right) \ , \quad \sigma^1 = \left( \begin{array}{cc}
0 & 1 \\
1 & 0 \end{array} \right) \ , \quad \sigma^2 = \left( \begin{array}{cc}
0 & -i \\
i & 0 \end{array} \right) \ ,  \quad \sigma^3 = \left( \begin{array}{cc}
1 & 0 \\
0 & -1 \end{array} \right) \ . \label{Pauli3}
\end{eqnarray}
Here $\sigma^0$ is the identity matrix and $\sigma^i$, $i=1,2,3$ are Pauli matrices. Let us suppose that we excite a gauge field of the following form
\begin{eqnarray}
A = A_0(t, r) \sigma^0 + A_3(t, r) \sigma^3 \ ,
\end{eqnarray}
where
\begin{eqnarray}
A^I (t, r) = \left(  - E_I t + a_x^I (r)  \right) dx \ , \quad  I = 0 , 3 \ . \label{nonAbeG}
\end{eqnarray}
This yields the following field-strength
\begin{eqnarray}
F_{\mu\nu} & = & F_{\mu\nu}^0 \sigma_0 + F_{\mu\nu}^3 \sigma_3 \nonumber\\
& = & \left( F_{tx}^0 \sigma_0 + F_{tx}^3 \sigma_3 \right) \delta_\mu^t  \delta_\nu^x + \left( F_{rx}^0 \sigma_0 + F_{rx}^3 \sigma_3 \right) \delta_\mu^r  \delta_\nu^x \nonumber \\
& = &  \left( - E^0 \sigma_0 - E^3 \sigma_3 \right) \delta_\mu^t  \delta_\nu^x + \left( \left(\partial_r a_{x}^0 \right) \sigma_0 +  \left(\partial_r a_{x}^3 \right)\sigma_3 \right) \delta_\mu^r  \delta_\nu^x \ .
\end{eqnarray}
Evidently, we have used the fact that $\sigma^{0,3}$ are diagonal and therefore commute with each other.

Now, to simplify the pull-back in (\ref{pbnonAbe}), we follow the arguments outlined in \cite{Erdmenger:2008yj}: First, using the $U(1) \subset U(2)$, we align the $\sigma^0$ flavour-direction parallel to the profile function $\Phi^8$. Subsequently, since we are exciting only an ``isospin" gauge direction (the one along $\sigma^3$), the representation in the flavour space is completely diagonal. This implies that all commutators in (\ref{pbnonAbe}) involving the embedding function $\Phi^8$ and the gauge field $A_\mu$ will vanish identically. Hence we completely decouple the fields $\Phi^8$ and $A_\mu$.

Thus, we evaluate the action in (\ref{nonAbe}) to be
\begin{eqnarray}
S_{\rm DBI} & = &  - 2 T_{{\rm D}7} \int d^8 \xi \sqrt{ - {\rm det} P[G]} \, \, {\rm STr} \left( \sigma^0 + || F ||^2 \right)^{1/2} \ , \\
|| F ||^2 & = & \left | \left | F_0 \sigma^0 + F_3 \sigma^3 \right | \right | ^2 \\
& = & G^{tt} G^{xx} \left( F_{tx}^0 \sigma_0 +  F_{tx}^3 \sigma_3 \right)^2 + G^{rr} G^{xx} \left( F_{rx}^0 \sigma_0 +  F_{rx}^3 \sigma_3 \right)^2 \nonumber\\
& = & G^{tt} G^{xx} \, {\rm diag} \left\{ \left( F_{tx}^0 + F_{tx}^3 \right)^2 , \left( F_{tx}^0 - F_{tx}^3 \right)^2 \right \} \nonumber\\
& + & G^{rr} G^{xx} \, {\rm diag} \left\{ \left( F_{rx}^0 + F_{rx}^3 \right)^2 , \left( F_{rx}^0 - F_{rx}^3 \right)^2 \right \} \ .
\end{eqnarray}
In the above calculations, we have used the explicit forms of the Pauli matrices given in (\ref{Pauli3}). Finally we get
\begin{eqnarray}
S_{\rm DBI} & = & - 2 T_{{\rm D}7} \int d^8 \xi \sqrt{ - {\rm det} P[G]} \left[ \left( 1 + G^{tt} G^{xx} E_+^2 + G^{xx} G^{rr} a_+'^2 \right)^{1/2} \right. \nonumber\\
& + & \left. \left( 1 + G^{tt} G^{xx} E_-^2 + G^{xx} G^{rr} a_-'^2 \right)^{1/2} \right] \ ,  \label{nAdbifinal}
\end{eqnarray}
with
\begin{eqnarray}
E_{\pm} = - \left( F_{tx}^0 \pm F_{tx}^ 3 \right) \ , \quad a_{\pm}(r) = a_x^0 (r) \pm a_x^3(r) \ ,
\end{eqnarray}
where $a^I$'s are first introduced in (\ref{nonAbeG}).

It is evident from (\ref{nAdbifinal}) that the embedding profile $\Phi$ is contained entirely inside $\left(\sqrt{ - {\rm det} P[G]}\right)$-piece of the action. Thus, if in the absence of $E_{\pm}$ and $a_{\pm}$, a profile function satisfies $ P[G] = G$, {\it i.e.}~$\Phi= {\rm constant}$, then at any finite value of $\{E_{\pm}, a_{\pm}\}$, $\Phi = {\rm constant}$ will also satisfy the equations of motion. This is what we normally call a {\it trivial embedding}. For now, we will restrict ourselves only to these trivial embeddings. It is also noteworthy that $E_{\pm}$ can be interpreted as independent electric fields at the boundary theory and $a_{\pm}$ should subsequently contain the information of the corresponding currents.

Now, the equation of motion resulting from (\ref{nAdbifinal}) are given by
\begin{eqnarray}
\frac{\sqrt{ - {\rm det} G} \, G^{xx} G^{rr} a_{\pm}'}{\left( 1 + G^{tt} G^{xx} E_{\pm}^2 + G^{xx} G^{rr} a_{\pm}'^2 \right)^{1/2}} = j_{\pm} \ ,
\end{eqnarray}
where $j_{\pm}$ are the corresponding constants of integration. Due to the decoupling of the $\pm$-modes, it is evident that there will be corresponding boundary currents 
\begin{eqnarray}
J_{\pm} \sim  \left. \frac{\delta S_{\rm DBI}}{\delta A_{\pm}} \right |_{\rm cut-off} \sim j_{\pm} \ .
\end{eqnarray}
Thus, classically, we are still describing a decoupled system driven with an external field $E_{\pm}$ and having a response current $j_{\pm}$.

Evaluated on an AdS-background, where $G_{tt} = - \frac{r^2}{L^2}$, $G_{xx} = \frac{r^2}{L^2} = G_{rr}^{-1}$, the solution for the decoupled gauge-modes are
\begin{eqnarray}
a_{\pm}' = \frac{j_{\pm}}{r^2} \frac{\sqrt{E_{\pm}^2 L^4 - r^4}}{\sqrt{j_{\pm}^2 - r^6 }}   \quad \implies \quad \lim_{r\to\infty} a_{\pm} = - \frac{j_{\pm}}{2 r^2} + \ldots \ .
\end{eqnarray}
Thus, the on-shell Lagrangian takes the following form
\begin{eqnarray}
\cL_{\rm DBI} \sim r^4 \left( \sqrt{\frac{E_{+}^2 L^4 - r^4} {j_{+}^2 - r^6}} + \sqrt{\frac{E_{-}^2 L^4 - r^4} {j_{-}^2 - r^6}} \right) \  .
\end{eqnarray}
Subsequently, we get two decoupled pseudo-horizons
\begin{eqnarray}
r_\ast^{(\pm)} = \left( E_{\pm}^2 L^4 \right)^{1/4} \ .
\end{eqnarray}
It is natural to expect that there will be two decoupled sets of fluctuations, corresponding to the $\pm$-modes, that will sense the above pseudo-horizon as the corresponding open string metric event-horizon.

Now, we discuss the physics of the fluctuation modes. To that end, let us imagine the following fluctuations\footnote{We will restrict ourselves to the scalar and the vector fluctuations only.}
\begin{eqnarray}
\Phi^{8,9} = \Phi_0^{8,9} + \delta \varphi^{8,9} \ , \quad A^a = A_{\rm class}^a + \delta A_0^a \sigma^0 + \delta A_i^a \sigma^i \ , 
\end{eqnarray}
where $\Phi_0^{8,9}$ describe the classical profile and $A_{\rm class}^0$ is given in (\ref{nonAbeG}). Thus, $\delta \varphi$ and $\delta A$ correspond to fluctuation modes. Evidently, we can write the fluctuation modes as
\begin{eqnarray}
\delta\varphi^i = \delta\varphi_I^i \, \sigma^I \quad \implies \quad \delta\varphi^i = \left( \begin{array}{cc}
\delta\varphi_+^i & \delta\varphi_{12}^i \\
\delta\varphi_{21}^i & \delta\varphi_-^i  \ \end{array} \right) \ ,
\end{eqnarray}
where
\begin{eqnarray}
\delta\varphi_{\pm} ^i = \delta\varphi_{0}^i \pm \delta\varphi_{3}^i \ , \quad \delta\varphi_{12}^i = \delta\varphi_{1}^i - i \delta\varphi_{2}^i \ , \quad \delta\varphi_{21}^i = \delta\varphi_{1}^i + i \delta\varphi_{2}^i \ .
\end{eqnarray}

As we have explained before, in this case the fluctuation action will be determined from the expansion
\begin{eqnarray}
S_{\rm fluc} & = & - 2 T_{{\rm D}7} \int d^8 \xi \, {\rm STr} \sqrt{ - {\rm det} a} \left(  1 + \frac{1}{2} {\rm tr}_{ab} \left( a^{-1} a^{(1)} \right)  + \frac{1}{8} \left({\rm tr}_{ab} \left( a^{-1} a^{(1)} \right) \right)^2 \right. \nonumber \\
& - & \left.  \frac{1}{4} {\rm tr}_{ab} \left( \left( a^{-1} a^{(1)} \right)^2 \right) + \frac{1}{2} {\rm tr}_{ab} \left( a^{-1} a^{(2)} \right) + \ldots \right) \ , \label{nonAkt}
\end{eqnarray}
where
\begin{eqnarray}
a_{ab} = G_{ab} + F_{ab} \ , \quad a_{ab}^{(1)} = \delta F_{ab}  \ , \quad a_{ab}^{(2)} = G_{ij} \partial_a \delta\varphi^i \partial_b \delta\varphi^j \ .
\end{eqnarray}
It is again clear that (\ref{nonAkt}) involves the inversion of the $a$-matrix in order to determine the ``effective metric" in the kinetic term for the fluctuations.

Now let us focus on the scalar kinetic term. Using (\ref{nonAkt}) and (\ref{osmsu2}) we get
\begin{eqnarray}
S_{\rm fluc}^{\rm scalar} & = & - 2 T_{{\rm D}7} \int d^8 \xi \sqrt{ - {\rm det} a} \, G_{ij} \left[ \cS_{ab}^{+} \left( \partial^a \delta\varphi_+^i \right) \left( \partial^b \delta\varphi_+^j \right) + \cS_{ab}^{-}  \left( \partial^a \delta\varphi_-^i \right) \left( \partial^b \delta\varphi_-^j \right) \right. \nonumber \\ 
 & + & \left.   \left( \cS_{ab}^{+} + \cS_{ab}^{-} \right) \left( \left( \partial^a \delta\varphi_{12}^i \right) \left( \partial^b \delta\varphi_{21}^j \right) + \left( \partial^b \delta\varphi_{12}^i \right) \left( \partial^a \delta\varphi_{21}^j \right) \right) \right]  + {\rm interactions} \ . \label{osmscalar}
\end{eqnarray}
Proceeding in a similar manner, we can also get
\begin{eqnarray}
S_{\rm fluc}^{\rm vector} & = & -  \frac{T_{{\rm D}7}}{2} \int d^8 \xi \sqrt{ - {\rm det} a} \left[  \cS^{+} _{ab} \cS^{+}_{cd} \,  \delta F_{+}^{ac} \delta F_{+}^{bd} + \cS^{-}_{ab} \cS^{-}_{cd} \, \delta F_{-}^{ac} \delta F_{-}^{bd} \right. \nonumber\\
& + & \left. \left(  \cS^{+}_{ab} \cS^{+}_{cd} + \cS^{-}_{ab} \cS^{-}_{cd} \right) \delta F_{12}^{ac} \delta F_{21}^{bd} \right] + {\rm interactions} \ . \label{osmvector}
\end{eqnarray}
The fluctuation action in (\ref{osmscalar}) and (\ref{osmvector}) are highly suggestive: Within each spin-sector, there are $3$ different modes (corresponding to ``$+$", ``$-$" and ``$12$") that sense three different effective geometries, subsequently given by $\cS^{+}$, $\cS^{-}$ and $\left(\cS^+ + \cS^- \right)$. These distinct modes ``see" three distinct event horizons, given by
\begin{eqnarray}
G_{tt} G_{xx} + \left( E_0 \pm E_3 \right)^2 = 0 \ , \quad {\rm and} \quad G_{tt} G_{xx} +  E_0^2 + E_3^2 = 0 \ . 
\end{eqnarray}
In AdS-space this yields
\begin{eqnarray}
r_\ast^{\pm} = \left( E_{\pm} ^2  \right)^{1/4} L   \ , \quad r_{\ast}^{\rm int} = \left[ \frac{1}{2}\left( E_{+}^2 - E_{-}^2 \right)\right]^{1/4} L \ .
\end{eqnarray}
Note that
\begin{eqnarray}
r_\ast^{+} > r_{\ast}^{\rm int}   \ . 
\end{eqnarray}
Clearly, $r_{\ast}^{\rm int} = 0 $, if $E_+ = E_-$, which means $F_{tx}^{3} = 0$.

\section{Conclusions}

In this article we have shown, with numerous examples and in details, how an open string equivalence principle emerges when one considers the fluctuations of open string degrees of freedom embedded in a closed string geometry. The examples we have discussed are mostly $10$-dimensional, {\it i.e.}~have well-defined stringy completions and hence well-defined dual field theories as well, at least in principle. This open string equivalence principle results in an effective temperature in the steady-state probe system. This effective temperature enters in {\it e.g.}~fluctuation-dissipation relations, among other things.

Many issues we have left untouched though. For example, we have completely ignored the backreaction of the probe sector. This backreaction will introduce explicit time-dependence in the problem since the constant energy that gets pumped into the system because of Ohmic dissipation, will eventually start increasing the background event horizon. If we keep the constant electric field turned on for a long but finite amount of time, it is likely that the dynamics will merge the background event horizon with the initial open string metric event horizon. This will be an interesting merger process to study. Furthermore, as we have seen, backreaction effects grow in the deep IR, which can lead to non-trivial ground states for the system.

Moving away from explicit time-dependence, the identification of the area of osm event horizon is an intriguing problem. It is also noteworthy that, the osm naturally allows a Kruskal extension and henceforth an Einstein-Rosen bridge. Although this is simply obtained by maximal analytic extension of a particular coordinate system, and does not come from an underlying dynamical process, this has the same flavour as the ER=EPR conjecture\cite{Maldacena:2013xja}, in the context of gauge-gravity duality\cite{Jensen:2013ora}. It will be extremely interesting to pin this problem down.

We end on the following note: As far as exploring governing principles of non-equilibrium physics is concerned, we have analyzed a tiny and a very special corner of it. It seems necessary that one studies the entire evolution process in details and look carefully for any universal qualitative feature that may survive the details of the system and the dynamics. Thus a complete understanding of the dynamical process is an enticing avenue for further and future work.

\section{Acknowledgements}

We thank Sumit Das, Jacques Distler, Willy Fischler, Vadim Kaplunovsky, Sandipan Kundu, Hong Liu, Gautam Mandal, Bala Sathiapalan and Julian Sonner for useful discussions and encouragements about this work. A special thanks go to Tameem Albash, Veselin Filev and Clifford Johnson for the wonderful collaborations that eventually led to this work. Furthermore we would like to thank Sandipan Kundu for collaborating on an earlier paper. A part of this work was done when AK was supported by a Simons postdoctoral fellowship awarded by the Simons Foundation. He would further like to thank the people of India for their generous support in research in basic sciences.

\renewcommand{\theequation}{A.\arabic{equation}}
\setcounter{equation}{0}  
\section*{Appendix A. Fluctuations of the classical profile}
\addcontentsline{toc}{section}{Appendix A. Fluctuations of the classical profile}

We will outline the steps towards obtaining action for the quadratic fluctuations. Without loss of generality, let us focus for a D$p$-brane, where $p \in [0,9]$. We will first discuss the bosonic part of the fluctuations. Let us denote the embedding of the probe brane by 
\begin{eqnarray} \label{flucx}
X^\mu = X_0^\mu + \Sigma^i \delta_i^\mu \ , 
\end{eqnarray}
where $X_0^\mu$ denotes the classical profile of the probe, and $\Sigma^i$ denotes the transverse fluctuations. Here the indices $\mu, \nu, \rho = 0, \ldots 9$; the indices $i, j, k = (p+1), \ldots , 9$ and the indices $a, b, c = 0 \ldots p$. Clearly, the Greek indices $\mu, \nu, \rho$ {\it etc.}are used for the ten dimensional background, the indices $i, j, k$ {\it etc.}~are used to represent the transverse fluctuations (which, for the D7-brane probes described in section \ref{n4sym} are denoted by $\delta\theta$ and $\delta\phi$ respectively), and finally the indices $a, b, c$ are used to represent the world volume of the probe brane.

The background metric $G_{\mu\nu}$ can be expanded under (\ref{flucx}) as
\begin{eqnarray}
G_{\mu\nu} = G_{\mu\nu}^0 + \left(\partial_\rho G_{\mu\nu}^0\right) \Sigma^i \delta_i^\rho + \frac{1}{2} \left(\partial_\rho\partial_\sigma G_{\mu\nu}^0\right) \Sigma^i \Sigma^j \delta_i^\rho \delta_j^\sigma + \ldots \ .
\end{eqnarray}
Here $G_{\mu\nu}^0$ denotes the background metric evaluated at the classical profile of the probe.

Now, the induced metric (up to quadratic order in the transverse fluctuations) on the probe D-brane can be obtained from the following formula
\begin{eqnarray}
g_{ab} & \equiv & P[G_{ab}] =  G_{\mu\nu} \partial_a X^\mu \partial_b X^\nu \nonumber\\
& = & g_{ab}^{(0)} + g_{ab}^{(1)} + g_{ab}^{(2)} \ ,
\end{eqnarray}
where
\begin{eqnarray} \label{gab0}
g_{ab}^{(0)} =  G_{\mu\nu}^0 \left(\partial_a X_0^\mu\right) \left( \partial_b X_0^\nu \right) \ ,
\end{eqnarray}
\begin{eqnarray} \label{gab1}
g_{ab}^{(1)} = \left(\partial_\rho G_{\mu\nu}^0 \right) \left(\partial_a X_0^\mu\right) \left(\partial_b X_0^\nu\right) \Sigma^i \delta_i^\rho + G_{\mu\nu}^0 \left[ \left(\partial_a X_0^\mu\right) \left(\partial_b\Sigma^i \right) \delta_i^\nu + \left(\partial_b X_0^\nu\right) \left(\partial_a\Sigma^i \right) \delta_i^\mu \right] \ ,
\end{eqnarray}
and 
\begin{eqnarray} \label{gab2}
g_{ab}^{(2)} & = & G_{\mu\nu}^0 \left(\partial_a \Sigma^i \right) \left(\partial_b \Sigma^j\right) \delta_i^\mu \delta_j^\nu + \frac{1}{2} \left(\partial_a X_0^\mu\right) \left(\partial_b X_0^\nu\right) \left(\partial_\rho \partial_\sigma G_{\mu\nu}^0 \right) \Sigma^i \Sigma^j \delta_i^\rho \delta_j^\sigma \nonumber\\
& + & \left(\partial_\rho G_{\mu\nu}^0\right) \Sigma^i \delta_i^\rho \left[ \left(\partial_a X_0^\mu \right) \left(\partial_b \Sigma^j \right) \delta_j^\nu + \left(\partial_b X_0^\nu \right) \left(\partial_a \Sigma^j \right) \delta_j^\mu \right] \ .
\end{eqnarray}
Now the action functional for the probe brane is given by
\begin{eqnarray}
S_{\rm probe}^{\rm boson} = - T_p  \int d^{p+1} \xi e^{- \Phi} \sqrt{ - {\rm det} \left(E_{ab}\right) } + \left({\rm Wess-Zumino}\right) \ ,
\end{eqnarray}
where $\xi$ represents the world volume coordinates, and the Wess-Zumino term depends on the dimensionality of the D$p$-brane as well as the background fluxes.

To proceed further, let us now define
\begin{eqnarray}
E_{ab} = E_{ab}^{(0)} + E_{ab}^{(1)} + E_{ab}^{(2)} \ , 
\end{eqnarray}
with
\begin{eqnarray}
E_{ab}^{(0)} = g_{ab}^{(0)} + f_{ab} + P[B] \ , \quad E_{ab}^{(1)} = g_{ab}^{(1)} + \delta f_{ab} \ , \quad E_{ab}^{(2)} = g_{ab}^{(2)} \ ,
\end{eqnarray}
where $P[B]$ is the pull back of the NS-NS $2$-form, $f_{ab}$ denotes the worldvolume gauge field and $\delta f_{ab}$ is the vector fluctuation. Now, to find the quadratic action, we use the following expansion
\begin{eqnarray}
{\rm det}(-E_{ab})^{1/2} = {\rm det}\left(-E_{ab}^{(0)} \right)^{1/2}  \left[ 1 + \frac{1}{2} {\rm Tr} M - \frac{1}{4} {\rm Tr} M^2 + \frac{1}{8} \left( {\rm Tr} M \right)^2\right] + \ldots \ ,
\end{eqnarray}
where
\begin{eqnarray}
M = \left(E^{(0)}\right)^{-1} \left(E^{(1)} + E^{(2)} \right) \ .
\end{eqnarray}
At the quadratic order, we get the following contributions
\begin{eqnarray}
&& \frac{1}{2} {\rm Tr} M \quad \implies \quad \frac{1}{2} {\rm Tr} \left[ \left(g^{(0)} + f \right)^{-1} g^{(2)}\right] \ , \\
&& \frac{1}{8} \left({\rm Tr} M\right)^2 \quad \implies \quad \frac{1}{8} {\rm Tr} \left[ \left(g^{(0)} + f \right)^{-1} \left(g^{(1)} + \delta f \right)\right] {\rm Tr} \left[ \left(g^{(0)} + f \right)^{-1} \left(g^{(1)} + \delta f \right)\right] \ , \\
&& \frac{1}{4} {\rm Tr} M^2 \quad \implies \quad \frac{1}{4} {\rm Tr} \left[ \left(g^{(0)} + f \right)^{-1} \left(g^{(1)} + \delta f \right) \left(g^{(0)} + f \right)^{-1} \left(g^{(1)} + \delta f \right) \right] \ .
\end{eqnarray}
It is evident that inverting the matrix $(g^{(0)} + f)$ is crucial in determining the quadratic action. Inversion of $(g^{(0)} + f)$ will consist of two pieces: a symmetric and an anti-symmetric one, denoted by $\cS$ and $\cA$ respectively. Following \cite{Seiberg:1999vs}, these pieces are obtained to be
\begin{eqnarray}
&& \left(\left(g^{(0)} + f \right)^{-1}\right)^{a b} = \cS^{ab} + \cA^{ab} \ , \label{defSA1} \\
&& \cS^{ab} = \left(\frac{1}{g^{(0)} + f } \cdot g^{(0)} \cdot \frac{1}{g^{(0)} - f }\right)^{ab} \ , \quad \cS_{ab} = g^{(0)}_{ab} - \left(f \cdot \left(g^{(0)}\right)^{-1} \cdot f \right)_{ab} \ , \label{defSA2}\\
&& \cA^{ab} = - \left(\frac{1}{g^{(0)} + f } \cdot f \cdot \frac{1}{g^{(0)} - f }\right)^{ab} \ . \label{defSA3}
\end{eqnarray}

Let us now discuss the fermionic part. The quadratic term is given by\cite{Marolf:2003ye, Marolf:2003vf, Martucci:2005rb}
\begin{eqnarray} \label{actferm}
S_{\rm probe}^{\rm fermion} = \frac{T_p}{2} \int d^{p+1}\xi e^{- \Phi} \sqrt{- {\rm det} \left(E^{(0)}\right)} \, \bar{\psi} \left(1 - \Gamma_p\right) \left( \Gamma^a D_a - \Delta + L_p \right) \psi \ ,
\end{eqnarray}
where $\psi$ is the fermionic fluctuation field. In type IIA, this is a $10$ dimensional Majorana spinor and in type IIB this is a positive chirality doublet Majorana-Weyl spinor. The worldvolume gamma matrices $\Gamma_a$ are pull-back of the $10$ dimensional gamma matrices $\Gamma_\mu$:
\begin{eqnarray}
\Gamma_a = \Gamma_\mu \left(\partial_a X^\mu \right) = \Gamma_{\underline{\mu}} e_{\mu} ^{\underline{\mu}} \left(\partial_a X^\mu\right) \ ,
\end{eqnarray}
where $X^\mu$ represents the $10$ dimensional spacetime, and the vielbein is given by
\begin{eqnarray}
e ^{\underline{\mu}} = e_{\nu} ^{\underline{\mu}} \, dX^\nu \ .
\end{eqnarray}
Let us now define the other operators appearing in (\ref{actferm}). For type IIA description (where $p= 2n$ is even-valued)
\begin{eqnarray}
&& \Gamma_{(2n)} = \sum_{q+r=n} \frac{(-1)^{r+1} \left(\Gamma_{(10)}\right)^{r+1} \epsilon^{a_1 \ldots a_{2q} b_1 \ldots b_{2r+1}}}{q! \left(2r + 1\right)! 2^q \sqrt{- {\rm det} E}} \cF_{a_1a_2} \ldots \cF_{a_{2q-1} a_{2q} } \Gamma_{b_1 \ldots b_{2r+1}} \ , \\
&&  L_{(2n)} = \sum_{q\ge 1, \, q+r=n} \frac{(-1)^{r+1} \left(\Gamma_{(10)}\right)^{r+1} \epsilon^{a_1 \ldots a_{2q} b_1 \ldots b_{2r+1}}}{q! \left(2r + 1\right)! 2^q \sqrt{- {\rm det} E}} \cF_{a_1a_2} \ldots \cF_{a_{2q-1} a_{2q} } \Gamma_{b_1 \ldots b_{2r+1}}^{\, \, \, \, \, \, \, \,  \, \, \, \, \, \, \, \, \, \, \, \, \, \, \, \, c} D_c  \ ,
\end{eqnarray}
whereas for type IIB (where $p = 2n+1$ is odd-valued) we have
\begin{eqnarray}
&& \Gamma_{(2n+1)} = \sum_{q+r=n+1} \frac{(-1)^{r+1} \left(i \sigma_2 \right) \left(\sigma_3 \right)^{r} \epsilon^{a_1 \ldots a_{2q} b_1 \ldots b_{2r}}}{q! \left(2r\right)! 2^q \sqrt{- {\rm det} E}} \cF_{a_1a_2} \ldots \cF_{a_{2q-1} a_{2q} } \Gamma_{b_1 \ldots b_{2r}} \ , \label{gammaiib} \\
&&  L_{(2n+1)} = \sum_{q\ge 1, \, q+r=n+1} \frac{(-1)^{r+1} \left(i \sigma_2 \right) \left(\sigma_3 \right)^{r} \epsilon^{a_1 \ldots a_{2q} b_1 \ldots b_{2r}}}{q! \left(2r\right)! 2^q \sqrt{- {\rm det} E}} \cF_{a_1a_2} \ldots \cF_{a_{2q-1} a_{2q} } \Gamma_{b_1 \ldots b_{2r}}^{\, \, \, \, \, \, \, \,  \, \, \, \, \, \, \, \, \, \,  c} D_c  \ .
\end{eqnarray}
Here $\sigma_{2,3}$ denotes the Pauli matrices, and $\epsilon$ is the epsilon-tensor. Furthermore, we have defined
\begin{eqnarray}
\cF = P[B] + f \ ,
\end{eqnarray}
where $P[B]$ denotes the pull back of the NS-NS anti-symmetric $2$-form and $F$ is the worldvolume gauge field. The operators $D_c$ and $\Delta$ are defined as follows:
\begin{eqnarray}
D_\mu = D_\mu ^{(0)} + W_\mu \ , \quad \Delta = \Delta^{(1)} + \Delta^{(2)} \ .
\end{eqnarray}
Now, in type IIA, we have
\begin{eqnarray}
D_\mu^{(0)} & = & \nabla_\mu + \frac{1}{4 \cdot 2!} H_{\mu\nu\rho} \Gamma^{\nu\rho} \Gamma_{(10)}  \ , \\
W_\mu & = & - \frac{1}{8} e^{\Phi} \left( \frac{1}{2} F_{\nu\rho} \Gamma^{\nu\rho} \Gamma_{(10)} + \frac{1}{4!} F_{\nu\rho\sigma\lambda} \Gamma^{\nu\rho\sigma\lambda} \right) \Gamma_\mu \ , \\
\Delta^{(1)} & = & \frac{1}{2} \left( \Gamma^\mu \partial_\mu \Phi + \frac{1}{2 \cdot 3!} H_{\mu\nu\rho} \Gamma^{\mu\nu\rho} \Gamma_{(10)} \right) \ , \\
\Delta^{(2)} & = & \frac{1}{8} e^{\Phi} \left( \frac{3}{2!} F_{\mu\nu} \Gamma^{\mu\nu} \Gamma_{(10)} - \frac{1}{4!} F_{\mu\nu\rho\sigma} \Gamma^{\mu\nu\rho\sigma} \right)  \ .
\end{eqnarray}
In type IIB, we have
\begin{eqnarray} 
D_\mu^{(0)} & = & \nabla_\mu + \frac{1}{4 \cdot 2!} H_{\mu\nu\rho} \Gamma^{\nu\rho} \sigma_3  \ , \\
W_\mu & = &  \frac{1}{8} e^{\Phi} \left( F_\mu \Gamma^\mu \left(i \sigma_2\right) + \frac{1}{3!} F_{\nu\rho\sigma} \Gamma^{\nu\rho\sigma} \sigma_1 + \frac{1}{2 \cdot 5!} F_{\nu\rho\sigma\lambda\kappa} \Gamma^{\nu\rho\sigma\lambda\kappa} \left(i \sigma_2\right)  \right) \Gamma_\mu \ ,  \label{wiib} \\
\Delta^{(1)} & = & \frac{1}{2} \left( \Gamma^\mu \partial_\mu \Phi + \frac{1}{2 \cdot 3!} H_{\mu\nu\rho} \Gamma^{\mu\nu\rho} \sigma_3 \right) \ , \\
\Delta^{(2)} & = & - \frac{1}{2} e^{\Phi} \left( F_\mu \Gamma^\mu \left(i \sigma_2\right) + \frac{1}{2 \cdot 3!} F_{\mu\nu\rho} \Gamma^{\mu\nu\rho} \sigma_1 \right)  \ ,
\end{eqnarray}
where 
\begin{eqnarray}
\nabla_\mu = \partial_\mu + \frac{1}{4} \Omega_{\mu} ^{\, \, \, \underline{\nu}\underline{\rho}} \Gamma_{\underline{\nu}\underline{\rho}}
\end{eqnarray}
is the covariant derivative.

Let us further note that, after making a few formal manipulations as outlined in \cite{Martucci:2005rb}, the operator $L_p$ can be rewritten as
\begin{eqnarray} \label{man1}
L_{(2n)} & = & - \Gamma_{(2n)} \sum_{q\ge 1} \left(\Gamma_{(10)}\right)^q \left( \cF^q\right)^{ab} \Gamma_a D_b \ , \\
L_{(2n+1)} & = & - \Gamma_{(2n+1)} \sum_{q\ge 1} \left(-\sigma_3 \right)^q \left( \cF^q\right)^{ab} \Gamma_a D_b \ ,
\end{eqnarray}
where we have defined
\begin{eqnarray}
\left( \cF^q \right)^{ab} = \cF^{a}_{\, \, \, \, \,  c_1} \cF^{c_1}_{\, \, \, \, \, c_2}  \ldots \cF^{c_{q-2}}_{\, \, \, \, \, \, \, \, \, \, \, \, \,  c_{q-1}} \cF^{c_{q-1} b} \ .
\end{eqnarray}
Furthermore, let us also define
\begin{eqnarray} \label{man2}
\tilde{E}_{ab}^{(0)} = g_{ab}^{(0)} + \tilde{\Gamma}_{(10)} \cF_{ab} \ ,
\end{eqnarray}
where
\begin{eqnarray} \label{man3}
{\rm type \, \, IIA}: \tilde{\Gamma}_{(10)} = \Gamma_{(10)} \ , \quad {\rm type \, \, IIB}: \tilde{\Gamma}_{(10)} = \Gamma_{(10)} \otimes \sigma_3 \ .
\end{eqnarray}
Using (\ref{man1})-(\ref{man3}), we can rewrite the action in (\ref{actferm}) as
\begin{eqnarray} \label{actferm2}
S_{\rm probe}^{\rm fermion} = \frac{T_p}{2} \int d^{p+1}\xi e^{- \Phi} \sqrt{- {\rm det} \left(E^{(0)}\right)} \, \bar{\psi} \left(1 - \Gamma_p\right) \left( \left(\tilde{E}_{(0)}^{-1}\right)^{ab} \Gamma_b D_a - \Delta  \right) \psi \ .
\end{eqnarray}
Thus the role of the operator $L_p$ is to redefine the kinetic term and we further observe that the inversion of the $\tilde{E}_{(0)}$ matrix, as defined in (\ref{man2}), plays a crucial role in determining the effective geometry observed by the fermionic fluctuations.

\renewcommand{\theequation}{B.\arabic{equation}}
\setcounter{equation}{0}  
\section*{Appendix B. Probe energy-momentum tensor}
\addcontentsline{toc}{section}{Appendix B. Probe energy-momentum tensor}

We can write the probe stress-energy in a covariant form. Imagine that the probe action is given by
\begin{eqnarray}
S_{\rm probe} = - \tau  \int [d\xi] e^{-\phi} \sqrt{- {\rm det} (G+F)} = - \tau \int [d\xi] e^{-\phi} \sqrt{- {\rm det} \cM}  \ .
\end{eqnarray}
Here $\tau$ is a overall constant that sets the brane tension. The variation of the above action will yield 
\begin{eqnarray}
 - \delta \sqrt{- {\rm det} \cM} \, e^{-\phi } = \frac{1}{2} e^{-\phi } \sqrt{- {\rm det} \cM} \,  \left( \cM^{ab} \delta \cM_{ab} \right) \ ,
\end{eqnarray}
where
\begin{eqnarray}
\cM^{-1} & = & \cS + \cA \ , \\
\cS^{ab} & = & \left( \frac{1}{G+F} \right)_{\rm sym}^{ab} = \left( \frac{1}{G+F} \cdot G \cdot \frac{1}{G-F}\right)^{ab} \ , \\
\cA^{ab} & = & \left( \frac{1}{G+F} \right)_{\rm anti-sym}^{ab} = - \left( \frac{1}{G+F} \cdot F \cdot \frac{1}{G-F}\right)^{ab} \ , \\
\delta \cM_{ab} & = & \delta G_{ab} + \delta F_{ab} \ .
\end{eqnarray}

Now, the variation of the probe action, evaluated on-shell, will yield the following stress-energy tensor
\begin{eqnarray}
| T_{tt} | & \sim & \kappa^2 T_{{\rm D}q} \frac{e^\phi}{2} \frac{G_c^{(p-m-9)/2}}{G_{xx}^{(p+1)/2}} \frac{J^2 E^2 e^{2\phi} G_c^{m+1} G_{xx} - |G_{tt}|^2 G_c^q G_{xx}^m G_{yy}}{\left[ \left( |G_{tt}| G_{yy} - E^2 \right) \left( |G_{tt}| G_c^q G_{xx}^m - J^2 e^{2\phi} G_c^{m+1} G_{xx} \right)\right]^{1/2}} \ , \label{Tprobe1} \\
| T_{yy} | & \sim & \kappa^2 T_{{\rm D}q} \frac{e^\phi}{2} \frac{G_c^{(p-m-9)/2}}{G_{xx}^{(p+1)/2}} G_{yy}^2 \sqrt{\frac{ |G_{tt}| G_{xx}^m G_c^q - J^2 e^{2\phi} G_c^{m+1} G_{xx}}{|G_{tt}| G_{yy} -E^2}} \ , \label{Tprobe2} \\
| T_{xx} | & \sim & \kappa^2 T_{{\rm D}q} \frac{(m-1)e^\phi}{2} \frac{G_c^{(p+2q-m-9)/2}}{G_{xx}^{(p-2m)/2}} \sqrt{\frac{G_{xx} (|G_{tt}| G_{yy} -E^2)}{|G_{tt}| G_{xx}^m G_c^q - J^2 e^{2\phi} G_c^{m+1} G_{xx}}} \ , \label{Tprobe3} \\
| T_{uu} | & \sim & \kappa^2 T_{{\rm D}q} \frac{e^\phi}{2} \frac{G_c^{(p -m-9)/2} G_{uu} }{|G_{tt}| G_{xx}^{(p+1)/2}} \left[ ( |G_{tt}| G_{yy} -E^2 ) (|G_{tt}| G_{xx}^m G_c^q - J^2 e^{2\phi} G_c^{m+1} G_{xx} ) \right ]^{1/2 } \ , \label{Tprobe4} \\
| T_{\alpha\beta} | & \sim & \kappa^2 T_{{\rm D}q} \frac{(m+1-q)e^\phi}{2} \frac{G_c^{(p+2q-m-7)/2}}{G_{xx}^{(p+1-2m)/2}} \sqrt{\frac{ (|G_{tt}| G_{yy} -E^2)}{|G_{tt}| G_{xx}^m G_c^q - J^2 e^{2\phi} G_c^{m+1} G_{xx}}} \eta_{\alpha\beta} \ . \label{Tprobe5}
\end{eqnarray}
Here $\eta_{\alpha\beta}$ represents the metric components on the unit-volume manifold $\cM_{8-p}$.

Here we will also derive the expression for the open string metric, both for the Abelian case and for the non-Abelian case. Let's consider the Abelian one first. Note the following:
\begin{eqnarray}
\left( G + F \right)^{-1} & = & \cS + \cA \ , \\
\left( G - F \right)^{-1} & = & \cS - \cA \ , 
\end{eqnarray}
which implies
\begin{eqnarray}
\cS & = & \frac{1}{2} \left[ \left( G + F \right)^{-1} + \left( G - F \right)^{-1} \right] \nonumber \\
& = & \left( G + F \right)^{-1} \cdot G \cdot \left( G - F \right)^{-1} \ . \label{symInv}
\end{eqnarray}
Similarly,
\begin{eqnarray}
\cA & = & \frac{1}{2} \left[ \left( G + F \right)^{-1} - \left( G - F \right)^{-1} \right] \nonumber \\
& = & - \left( G + F \right)^{-1} \cdot F \cdot \left( G - F \right)^{-1} \ .
\end{eqnarray}
From (\ref{symInv}), it is clear that 
\begin{eqnarray}
\cS^{-1} = \left( G - F \right) G^{-1} \left( G + F \right) = G - \left( F G^{-1} F\right) \ . \label{symInvlower}
\end{eqnarray}
Clearly, (\ref{symInv}) and (\ref{symInvlower}) define the open string metric with upper and lower indices respectively.

Let us now consider the case of $SU(2)$. In this case, we have defined
\begin{eqnarray}
a = G \sigma^0 + F_0 \sigma^0 + F_3 \sigma^3 = \left( \begin{array}{cc}
G + F_0 + F_3  & 0 \\
0 & G + F_0 - F_3  \end{array} \right)  \ . 
\end{eqnarray}
Proceeding along similar lines as outlined above, we can show
\begin{eqnarray}
\cS & = & \left ( \left(G + F_0 \right) \sigma^0 + F_3 \sigma^3 \right)^{-1} \cdot G \cdot \left(  \left(G - F_0 \right) \sigma^0 - F_3 \sigma^3 \right)^{-1} \ . \\
\implies \cS^{-1} & = & \left[ G - F_0  G^{-1}  F_0 - F_3  G^{-1}  F_3 \right] \sigma^0 + \left( F_0 G^{-1} F_3 - F_3 G^{-1} F_0 \right) \sigma^3  \label{osmsu2} 
\end{eqnarray}
To derive the last line, we have used the fact $\left(\sigma^3 \right)^2 = \sigma^0$ and $\sigma^3 \sigma^0 = \sigma^0 \sigma^3 = \sigma^3$. We can write down the open string metric (with indices lowered) as
\begin{eqnarray}
\cS^{-1} & = & \left( \begin{array}{cc}
\cS^{+} & 0 \\
0 & \cS^{-}  \end{array} \right)  \ , \nonumber \\ 
\cS^{\pm} & = & G - \left( F_0 \pm F_3 \right) G^{-1} \left( F_0 \pm F_3 \right)  \ . 
\end{eqnarray}
%



\end{document}